\documentclass[modern]{aastex62}

\usepackage{natbib}
\usepackage[utf8]{inputenc}
\usepackage{amsmath,amsfonts}
\usepackage{bbold}
\usepackage{graphicx}
\usepackage{xspace}

\graphicspath{{./}{figs/}}

\hypersetup{
    colorlinks=true,
    linkcolor=blue,
    filecolor=magenta,      
    urlcolor=cyan,
    pdftitle={Bolometric treatment of irradiation effects},
    pdfpagemode=FullScreen
    }

\shorttitle{Bolometric treatment of irradiation effects}

\shortauthors{Horvat et. al.}

\newcommand{\phoebe}{PHOEBE\xspace}

\begin{document}

\title{Bolometric treatment of irradiation effects:\\ general discussion and application to binary stars}

\author[0000-0002-0504-6003]{Martin Horvat}
\affiliation{University of Ljubljana, Dept.~of Physics, Jadranska 19, SI-1000 Ljubljana, Slovenia}
\affiliation{Villanova University, Dept.~of Astrophysics and Planetary Sciences, 800 E Lancaster Ave, Villanova PA 19085, USA}

\author[0000-0002-5442-8550]{Kyle E.~Conroy}
\affiliation{Villanova University, Dept.~of Astrophysics and Planetary Sciences, 800 E Lancaster Ave, Villanova PA 19085, USA}

\author[0000-0003-3947-5946]{David Jones}
\affiliation{Insituto de Astrof\'isica de Canarias, E-38205 La Laguna, Tenerife, Spain}
\affiliation{Departamento de Astrof\'isica, Universidad de La Laguna, E-38206 La Laguna, Tenerife, Spain}

\author[0000-0002-1913-0281]{Andrej Pr\v sa}
\affiliation{Villanova University, Dept.~of Astrophysics and Planetary Sciences, 800 E Lancaster Ave, Villanova PA 19085, USA}

\email{martin.horvat@fmf.uni-lj.si}

\begin{abstract}
A general framework for dealing with irradiation effects in the bolometric sense --- specifically, reflection with heat absorption and the consequent redistribution of the absorbed heat --- for systems of astrophysical bodies where the boundaries are used as support for the description of the processes, is presented. Discussed are its mathematical and physical properties, as well as its implementation approximations, with a focus on three plausible redistribution processes (uniform, latitudinal, and local redistribution). These are tested by extending PHOEBE 2.1 (\url{http://phoebe-project.org/}), the open-source package for modeling eclipsing binaries, and applied to a toy model of the known two-body eclipsing systems.
\end{abstract}


\keywords{Methods: analytical, numerical; Techniques: photometric, Binaries: eclipsing; Stars: fundamental parameters}

\section{Introduction}

There are at least three fundamentally different approaches for dealing with the reflection effect in binary stars. The most precise, and typically most time-consuming, approach is to treat the stellar atmospheres in detail, as static or given by hydrodynamics,  and use radiative transfer to calculate new temperatures and fluxes emitted from each of the stars  \citep[see, e.g.,][and references therein]{nordlund1990,hubeny2003,dobbs-dixon2013} .  A much simpler, but less precise, treatment can be achieved by using standard (non-irradiated) model atmospheres to approximate the flux emitted from the individual stars, which is then reflected between surfaces multiple times, thereby effectively heating the surfaces. The most widely spread example of such an approach is the Wilson reflection  model \citep{wilson1990}. For a very illustrative description of the latter, see, e.g., \cite{kallrath2009}. A similar study of reflection effects in binary systems was conducted by \cite{mochnacki1972} and \cite{hendry1992}, where geometrical aspects of multiple mutual irradiation in the standard Roche model was developed. However, under this scheme, energy is not always conserved, with some fraction of the incident flux being reflected while the rest is essentially ignored. The first model that addresses this issue, combining reflection with the redistribution of absorbed energy across the surfaces, is presented by \citet{budaj2011} and implemented in the SHELLSPEC code \citep{budaj2004}. Their model focuses on an effective description of uniform and latitudinal redistribution in Roche geometry. Finally, the fastest and arguably least informative methodology for approximating irradiation is based on relative corrections of the observables due to reflection. Such an approach was used in \cite{barclay2012} and \cite{barclay2015}, where they take an analytic model for light curves (LCs) in binary systems of spherical bodies with quadratic limb darkening, provided by the \cite{mandel2002} transit model and account for reflection by correcting the observed flux by a simple and analytic phase-dependent multiplicative factor.

We present a consistent mathematical model for handling the reflection effect from astrophysical bodies with redistribution of the absorbed irradiation in the directions laid out by \cite{budaj2011}.  We refer to these combined effects as irradiation effects. We consider only stationary or quasi-stationary redistribution, where we assume that the energy balance is fulfilled at all times, but the redistribution rules may change in time.  If these changes are slow in comparison to the flux transport, the quasi-stationary assumption is physically justified. In addition, we assume that the redistribution does not significantly change the limb darkening law of the considered surface \citep[a necessary assumption given that redefining the limb darkening would require detailed treatment of irradiation in the stellar atmosphere models;][]{claret2007}.  Our discussion is limited to the purely bolometric treatments of irradiation effects with the (Bond)\footnote{Here the albedo is assumed to lie in the range $[0,1]$ and is sometimes called the Bond albedo to differentiate it from the geometric albedo.} bolometric albedo  depending on the position on the surface. Nevertheless, we acknowledge that such a bolometric albedo is a poor representation of the effective albedo, which is essentially wavelength and temperature dependent \citep{vaz1985}. The bolometric albedo is generally assumed to be non-unity, which is discussed in \citet{rucinski1969} for different types of stellar envelopes. As highlighted above, the energy balance in several standard reflection models (e.g.,  \citealt{wilson1990}, \citealt{prsa2005}) is violated when the bolometric albedo deviates from unity.  An approximate passband dependence of synthetic observations can be obtained using  \citeauthor{wilson1990}'s spectral re-interpretation of bolometric results. A rigorous (i.e. approximation-less) passband-dependent treatment of reflection remains an unsolved problem.

The reflection--redistribution (or simply irradiation) framework presented here is tested by extending the publicly accessible Python package \phoebe available at \url{http://phoebe-project.org/}, which internally handles limb darkening, gravity brightening, Doppler shifts, and other details that determine emission properties of the considered astrophysical bodies. As the bolometric process discussed here is quite limiting for applications, this paper is not accompanied with a new release of the code.  If a passband treatment is developed in the future, the code will be released at that time.  This work can be considered a generalization of reflection--redistribution model introduced by \cite{budaj2011}, making it more geometry independent and easily extendable with different redistribution types.

We start the paper by introducing some common notation to describe the irradiation processes which are then used to define different reflection schemes and redistribution effects. Our description relies heavily on linear operators that make expressions more compact and readable, and compatible with modern implementations. We then explain the discretization of the introduced operators on triangular surfaces and outline practical considerations relating to the implementation of irradiation schemes. We conclude the paper with demonstrations of these principles on a toy model binary system.

\section{Description of bolometric irradiation}

Each body forms a closed boundary, ${\cal M}_i$. The union of all such boundaries constitutes the \emph{topological surface} ${\cal M} = \bigcup_i {\cal M}_i$. At each point ${\bf r}$ on the surface $\cal M$, we have a normal vector $\hat {\bf n}({\bf r})$ pointing outward from the body's interior. We work strictly with bolometric quantities. This restriction simplifies our discussion. Let us define a visibility function $V({\bf r}, {\bf r}')$ between the two points ${\bf r}$, ${\bf r}' \in \cal M$ as
\begin{equation}    
  V({\bf r}, {\bf r}') = \left \{
  \begin{array}{lll}
  1&:& \textrm{line of sight } {\bf r}  \leftrightarrow {\bf r}' \textrm{ is unobstructed}\\
  0&:& \textrm{otherwise}
  \end{array}\right. \>.
\end{equation}
In a system of two convex bodies, this visibility function is given by
\begin{equation}
    V({\bf r}, {\bf r}') = 
    U\left(
        \hat {\bf e}({\bf r}, {\bf r}')\cdot \hat {\bf n}({\bf r})
    \right)
    U\left(
    \hat {\bf e}({\bf r}', {\bf r})\cdot \hat {\bf n}({\bf r'})
    \right)\>,
\end{equation}
where $U(x)=\{1\mathpunct{:}~x\ge 0; 0\mathpunct{:}~{\rm otherwise}\}$ is the step-function and $\hat {\bf e}({\bf r}, {\bf r}') = \widehat{{\bf r}-{\bf r}'}$ denotes the unit vector pointing from ${\bf r}'$ to ${\bf r}$.

In order to facilitate the discussion that follows, we start with a concise glossary of the frequently used radiometric terms, and we direct the reader for further details to  \citet{modest2003} and \citet{hapke2012}.

\paragraph{Intensity} the energy flux $\Phi$ in the direction $\hat {\bf e}(\|\hat{\bf e}\|=1)$ from point ${\bf r}$ on the surface per solid angle per unit area normal to the surface (projection unit area) . It is here denoted by $I(\hat{\bf e}, {\bf r})$ and given by expression
\begin{equation}    
 I(\hat{\bf e}, {\bf r}) = 
 \frac{{\rm d}^2 \Phi}{{\rm d} \Omega {\rm d} A \cos \theta}\>,
 \qquad
 \cos\theta =\hat{\bf e} \cdot \hat{\bf n}({\bf r}) \>,
\end{equation}
where ${\rm d}\Omega$ is differential of the solid angle and  ${\rm d}A \cos\theta$ the differential of the surface area perpendicular to the normal. In modern radiometry literature, this quantity is usually called the radiance, and the intensity is then defined as the surface integral of the radiance. It is also referred to as bolometric intensity or total intensity in \citet{modest2003}.

\paragraph{Irradiance} radiant flux per unit area intercepted by a surface at a certain point ${\bf r}$, denoted by $F_{\rm in}({\bf r})$, with the index indicating that the energy flux is directed toward the surface. It is a non-directional quantity. If the surface intensity is $I$, then the irradiance is defined as 
\begin{align}
 F_{\rm in} ({\bf r}) 
  &= \int_{\cal M} V({\bf r}, {\bf r}')
  \frac{
  ({\hat {\bf e} ({\bf r}', {\bf r}) } \cdot \hat {\bf n}({\bf r}))
  ({\hat {\bf e}}({\bf r}, {\bf r}') \cdot \hat {\bf n}({\bf r'}))
  }{|{\bf r} - {\bf r}'|^2} 
  I({\hat {\bf e}}({\bf r}, {\bf r}'), {\bf r}') {\rm d} A ({\bf r'})\\
  &\equiv \hat {\cal Q} I ({\bf r}). \label{eq:reflection_R}
\end{align}
For simplicity, we introduce the irradiation operator $\hat {\cal Q}$ for mapping intensities to irradiances.

\paragraph{Radiant exitance} energy flux emitted at a certain point ${\bf r}$ on a surface per unit area. It is a non-directional quantity, and it does not include any reflected flux. In a general context, it is denoted by $F_{\rm ext}$, but if we talk about intrinsic exitance and updated intrinsic exitance, these are denoted by $F_0({\bf r})$ and $F_0'({\bf r})$, respectively. Radiant exitance is frequently referred to only as exitance. Here we discuss two different functional forms of the intensity $I$ deduced from radiant exitance:

\begin{itemize}
    \item[(a)] For a surface behaving as a Lambertian radiator, the intensity described by the Lambert cosine law \citep[Ch.~8.5.1]{hapke2012} is
\begin{equation}    
  I_{\rm L}(\hat {\bf e}, {\bf r}) = I_0({\bf r}) \>.
\end{equation}
The resulting radiant exitance is given by
\begin{equation}
 F_{\rm ext, L}({\bf r}) = \int_{\hat {\bf n}\cdot\hat {\bf e}\ge 0} 
 I_{\rm L}(\hat {\bf e}, {\bf r}) \, (\hat{\bf e} \cdot \hat {\bf n}) \, {\rm d} \Omega(\hat {\bf e})
 = \pi I_0({\bf r}) \>.
\end{equation}
\item[(b)] Typically, the light emission from the surface is described by limb-darkened intensity \citep{wilson1990}
\begin{equation}  
  I_{\rm LD}(\hat {\bf e}, {\bf r}) = I_0({\bf r}) 
  D(\hat {\bf e}\cdot \hat {\bf n}, {\bf r} ) \>,
  \label{eq:intesities}
\end{equation}
where $I_0({\bf r})$ is the normal emergent intensity and $D$ is the limb darkening factor; $D(1, {\bf r}) =1$. The corresponding radiant exitance then becomes
\begin{equation}    
  F_{\rm ext, LD}({\bf r}) = 
  \int_{\hat {\bf n}\cdot\hat {\bf e}\ge 0}
  I_{\rm LD} (\hat{\bf e}, {\bf r}) 
  \, (\hat{\bf e}\cdot \hat {\bf n})\, {\rm d} \Omega(\hat {\bf e})
  =  I_0({\bf r}) D_0({\bf r})\>,
\end{equation}
with the integrated limb darkening factor over the hemisphere
\begin{equation}
      D_0({\bf r}) = \int_{\hat {\bf n}\cdot\hat {\bf e}\ge 0}
  D(\hat{\bf e}\cdot \hat {\bf n}, {\bf r} ) \,( \hat{\bf e}\cdot \hat {\bf n} ) \, {\rm d} \Omega(\hat {\bf e})\>.
\end{equation}
\end{itemize}

\paragraph{Reflection} loosely a process by which a part of the energy flux received by the surface from outside is emitted (transmitted) back into space, depending on the reflection model. If the surface is treated as an ideal Lambertian radiator, we can say that the light is transmitted from the surface as its interior is not participating in the process. On the other hand,  the use of limb darkening in the \citeauthor{wilson1990}'s reflection model indicates that the reflection in this model is could be considered as (re-)emission of received flux from the atmosphere. As we approach the problem in a purely bolometric sense, we assume that reflectance is wavelength independent as in \cite{wilson1990} and \cite{budaj2011}, and consequently, the Bond albedo is identical to bolometric albedo, denoted here by $\rho$, and represents the fraction of incoming flux that is reflected. For more information on radiometric measures of reflection, see, e.g., \citet[Ch.~11.3]{hapke2012}.

\paragraph{Radiosity} energy flux per unit area that leaves (is emitted, reflected, and transmitted) a certain point ${\bf r}$ on the surface. It is a non-directional quantity. It is denoted by $F_{\rm out}({\bf r})$, with the index indicating that the energy flux is directed away from the surface. In our work, the radiosity is a sum of exitance (intrinsic or updated intrinsic) and reflected irradiance,
\begin{equation}
    F_{\rm out}({\bf r}) = F_{\rm ext}({\bf r}) + \rho({\bf r}) F_{\rm in}({\bf r}),
\end{equation}
where $\rho({\bf r})$ is local bolometric albedo at point ${\bf r}$ on the surface.

\begin{figure}[!htb]
\centering
\includegraphics[width=0.7\textwidth]{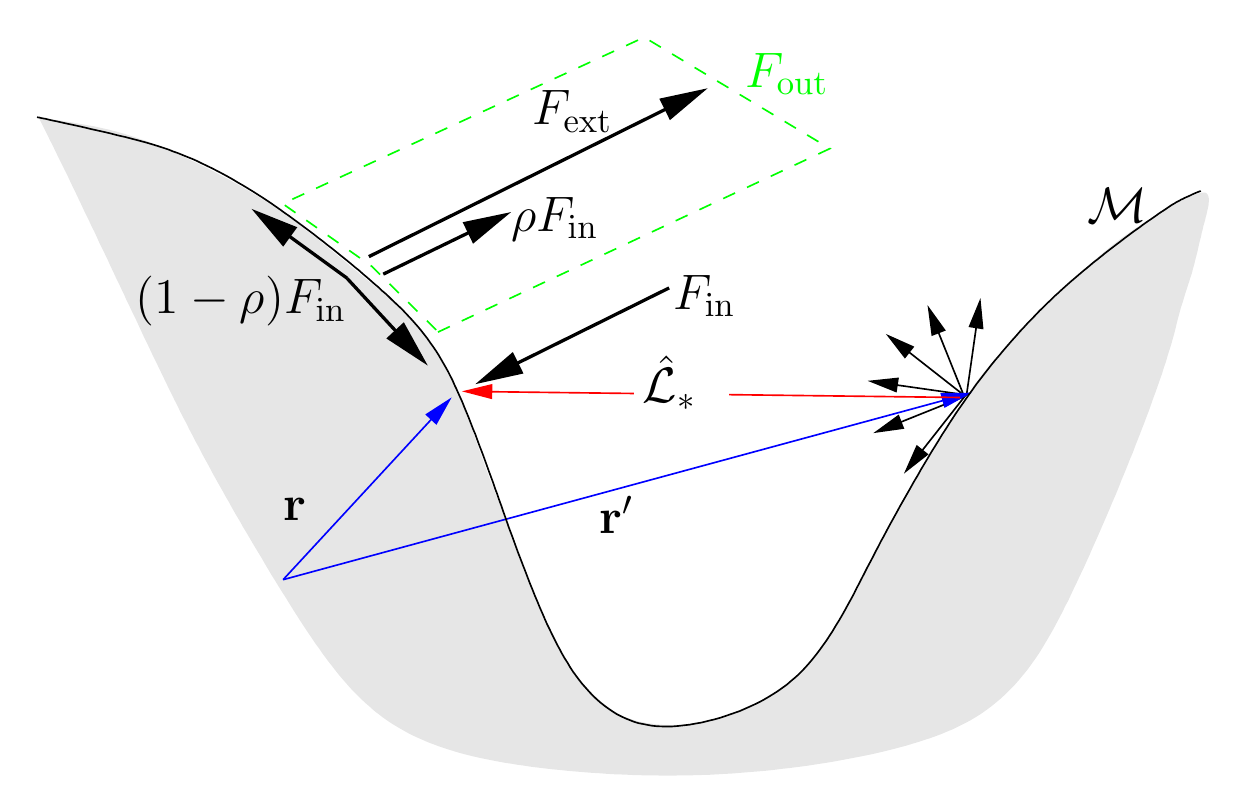}
\caption{Illustration of flux emitted from the surface $F_{\rm ext}$ and irradiation $F_{\rm in}$ calculated through the radiosity operators $\hat{ {\cal L}}_*$, with $*=\textrm{L or LD}$ and the process of reflection; where at point $\bf r$ the reflected part of the irradiation is $\rho({\bf r}) F_{\rm in}({\bf r})$ and the absorbed part is  $(1-\rho({\bf r})) F_{\rm in}({\bf r})$. The radiosity $F_{\rm out}$ is a sum of the exitance and the reflected part of the irradiation.}
\label{fig:illustration}
\end{figure}

\paragraph{Redistribution} the process by which the absorbed energy flux at the surface, i.e., the part of the flux that is not reflected, is redistributed across the surface. This additional energy flux is subsequently re-emitted from the surface. We assume that the diffusion of energy from the surface obeys the same limb darkening law with or without redistribution. This can only be valid if structural changes of atmosphere due to redistribution are relatively small. \\[2mm]
The introduced irradiation processes, composed of reflection and redistribution, are schematically depicted in Fig.~\ref{fig:illustration}. In the context of the paper, it is convenient to express the irradiance $F_{\rm in}$ as a consequence of the exitance $F_{\rm ext}$. In the case of a Lambertian surface, the irradiance $F_{\rm in, L}$ is connected to the exitance $F_{\rm ext, L}$ by introducing a Lambertian \emph{radiosity} operator $\hat {\cal L}_{\rm L}$:
\begin{equation}
  F_{\rm in,L} = \hat {\cal L}_{\rm L} F_{\rm ext, L}
  \qquad 
  \hat {\cal L}_{\rm L} = \frac{1}{\pi} \hat {\cal Q} \>.
  \label{eq:reflection_L_L}
\end{equation}
In the presence of limb darkening, the irradiance $F_{\rm in, LD}$ is expressed via the exitance $F_{\rm ext, L}$ and the limb-darkened radiosity operator $\hat {\cal L}_{\rm LD}$:
\begin{equation}
  F_{\rm in, LD} = \hat {\cal L}_{\rm LD} F_{\rm ext, LD}
  \qquad 
  \hat {\cal L}_{\rm LD} = \hat {\cal Q} \circ \frac{D}{D_0} \>.
  \label{eq:reflection_L_LD}
\end{equation}
The radiosity operators $\hat {\cal L}_{\rm L}$ and $\hat {\cal L}_{\rm LD}$ are an elegant mathematical way of expressing the relation between the two non-directional quantities, i.e., irradiation and exitance. Radiosity operators are commonly used in computer graphics (see \citealt{gershbein1994, cohen2016}).

\section{Reflection models}

Let us assume that we know the intrinsic exitance $F_0$ and the reflection fraction $\rho$ (i.e.~the ratio of the incoming energy flux per unit area that is reflected) for each point on the surface $\cal M$. The intrinsic exitance irradiates the unobstructed surface, some of which reflects back and irradiates the radiating surface, iteratively. We quantify the resulting radiosity (i.e. the radiant flux leaving the surface per unit area) $F_{\rm out}$ by introducing a reflection model.

We discuss two reflection models in detail: that proposed by \citet{wilson1990} and the Lambertian model introduced by \citet{prsa2016}. \citeauthor{wilson1990}'s model is based on the following set of equations:
\begin{equation}
  F_{\rm in} = \hat {\cal L}_{\rm LD} F_{\rm out} \qquad
  F_{\rm out} = F_0  + \hat \Pi F_{\rm in} \>,
  \label{eq:wilson_reflection}
\end{equation}
while \citeauthor{prsa2016}'s reflection model is based on the following set:
\begin{equation}
  F_{\rm in} = \hat {\cal L}_{\rm LD} F_{\rm 0} + \hat {\cal L}_{\rm L} \hat \Pi F_{\rm in}\qquad
  F_{\rm out} = F_0  + \hat \Pi F_{\rm in} \>,
  \label{eq:phoebe_reflection}
\end{equation}
where $\hat {\cal L}_{\rm LD}$ and $\hat {\cal L}_{\rm L}$ are the radiosity operators introduced by Eqs.~(\ref{eq:reflection_L_LD}) and (\ref{eq:reflection_L_L}), respectively. Additionally, we introduce the reflection operator $\hat \Pi:f \mapsto \rho f$, with $\rho$ being a scalar function defined on the surface describing the local bolometric albedo, i.e., the fraction of reflected light for each point separately.
 
The set of Eqs.~(\ref{eq:phoebe_reflection}) can be combined into a single expression for radiosity $F_{\rm out}$:
 \begin{equation}
  F_{\rm out} = \left[\mathbb{1} + \hat \Pi (\hat {\cal L}_{\rm LD} - \hat {\cal L}_{\rm L})\right] F_{\rm 0}  + \hat \Pi \hat {\cal L}_{\rm L} F_{\rm out} \>.
  \label{eq:phoebe_reflection2}
\end{equation}
It is evident from this expression that \citeauthor{prsa2016}'s model reduces to \citeauthor{wilson1990}'s in the limit $\hat {\cal L}_{\rm LD} = \hat {\cal L}_{\rm L}$. In \citeauthor{wilson1990}'s model, the radiation from the surface is distributed as a limb-darkened intensity, while in \citeauthor{prsa2016}'s approach only the intrinsic part of the radiance is distributed according to the limb-darkened intensity, while the reflected irradiance is distributed according to the Lambertian cosine law. Appendix \ref{sec:integral_form} provides these equations in integral form for the case of two convex radiators.

These reflection models do not conserve energy because the absorbed part of the irradiance, $(1-\rho) F_{\rm in}$, is dropped from the energy balance. When fitting models to the data, this energy loss can be compensated by increasing the albedo or the effective temperature of the radiators. Flux conservation is systematically corrected by the redistribution processes.

\section{Stationary and quasi-stationary redistribution}

The redistribution of the incoming energy flux depends on the thermodynamical circumstances in stellar photospheres, i.e., mechanical flows of matter and lateral thermal gradients, which can be very complex. However, for as long as these circumstances are long-lived insofar that we can assume stationary or quasi-stationary equilibrium for each star, we can build a framework to describe the redistribution of energy. We present such a framework and provide approximations for several simple scenarios. We define redistribution as a linear mapping of the incident flux density onto the radiated flux density, both defined on the surface of the body. The linearity assumption implies that the redistribution processes are independent from the flux scaling factors.   

\subsection{Lossless reflection--redistribution models}

We are primarily focusing on radiating bodies with a simple geometry, especially close to spherical, where we can describe the redistribution processes and identify the surface parts associated with flux incidence and emission. For strongly deformed bodies, such as contact binaries, we currently lack sufficient insight to propose a realistic yet tractable model because, as of yet, the thermodynamical properties are not well understood. Substantial effort to better understand radiative transfer in strongly deformed bodies is underway \citep{kochoska2019}.

In the stationary or quasi-stationary state, we assume that the flux is strictly conserved at all times, meaning that the net incident flux is also emitted at the same time. The absorbed part of the irradiance, $(1-\rho) F_{\rm in}$, is redistributed over the entire surface, and it increases the intrinsic exitance by $\delta F_0$. We can describe flux redistribution by defining a redistribution operator $\hat{\cal D}$,
\begin{equation}
  \delta F_0 
    = \hat {\cal D}\left[(1-\rho) F_{\rm in}\right] 
    = \hat {\cal D}  (\mathbb{1}-\hat \Pi) F_{\rm in} \>.
  \label{eq:delta_intrinsic}
\end{equation}
This redistribution operator could in principle be time-dependent in the quasi-stationary case, but for the strictly stationary state redistribution, the operator does not depend on time. The increased intrinsic exitance $F_0'$ can be written as
\begin{equation}
  F_0' = F_0 + \delta F_0 \qquad F_{\rm out} = F_0' + \hat \Pi F_{\rm in} \>.
  \label{eq:new_exitance}
\end{equation}
where we work under the assumption that the intrinsic exitance $F_0$ is not affected by irradiation. By construction, the redistribution operator $\hat{\cal D}$ maps positive-valued functions defined over the surface to positive-valued functions and conserves their integrals over the surface.

Generally, we can write the redistribution operator $\hat{\cal D}$ at time $t$ as:
$$
  \hat {\cal D} f({\bf r}) = 
  \int_{\cal M} K({\bf r},{\bf r'};t) f({\bf r'}) \, {\rm d} A({\bf r'})\>,
$$
where $K:{\cal M} \times {\cal M} \to \mathbb{R}_+$ is a positive kernel with the following normalization property:
$$
  \int_{\cal M} K({\bf r},{\bf r'};t) \, {\rm d} A({\bf r}) = 1\qquad \forall {\bf r} \in {\cal M} \>.
$$
This imposes the conservation of total flux at a given moment in time over the surface:
\begin{equation}  
  \int_{\cal M} [\hat {\cal D}f]({\bf r}) \, {\rm d} A({\bf r})
  = \int_{\cal M} f({\bf r}) \, {\rm d} A({\bf r}) \>.
  \label{eq:redistr_op_property}
\end{equation}
More generally, we can formulate the kernel by using an auxiliary function $G:{\cal M} \times {\cal M} \to \mathbb{R}_+$:
$$
  K({\bf r}, {\bf r}';t) = 
    \frac
        {G({\bf r},{\bf r}';t)}
        {\int_{\cal M} G({\bf r},{\bf r'};t) {\rm d} A({\bf r}) } \>.
$$
When the flux is uniformly distributed over the whole surface, $G_{\rm uniform} \equiv 1$ and the redistribution operator $\hat{\cal D_{\rm uni}}$ is very simple:
$$
  \hat {\cal D}_{\rm uni} f({\bf r})  
  = \frac{1}{A}\int_{\cal M} f({\bf r'}) {\rm d} A ({\bf r'}) \>.
$$

We model a local redistribution by introducing a distance measure (e.g.~a geodesic) on the surface, $d:{\cal M}\times{\cal M} \to \mathbb{R}_+$, and a weight function, $g: \mathbb{R}_+ \to \mathbb{R}_+$, that determines the ratio of the flux that is transported from the irradiated element to any other element on the surface at distance $d$. Then, the kernel describing the redistribution of incident flux is written as
\begin{equation}    
  G_{\textrm{loc}}({\bf r}, {\bf r}';t) = g\left(d({\bf r}, {\bf r'})\right) \>.
  \label{eq:Glocal}
\end{equation}
The redistribution operator associated with this kernel is denoted by $\hat {\cal D}_\textrm{loc}$. The weight function $g$ depends on the energy transport in the atmosphere, which we do not (readily) know. However, we can make reasonable assumptions that are likely to hold. Namely, we take $g$ to be monotonically decreasing and diminishing to zero for arguments larger than a given threshold value $l$, for example, $g(x) = \exp(-x/l)$ or $g(x) = 1-x/l$, where $l$ is proportional to the optical depth in the atmosphere. In the limit $l\to 0$, local redistribution reflects all incoming flux: $\hat {\cal D}_{\rm loc}= \mathbb{1}$ at $l=0$. Therefore, local redistribution with small $l$ (w.r.t.~to the size of object) will have an effect similar to increasing reflection. For spherical bodies of radius $R$, it is much more convenient to use the ratio $l/R$ as a parameter determining the threshold value.

In the case of rotating stars, the axis of rotation breaks the isotropic symmetry, so excess flux tends to be reradiated at latitudes similar to where it was received. This gives rise to flux conservation in the latitudinal direction. It is meaningful to define a distance on the surface, $d_\perp:{\cal M}\times{\cal M} \to \mathbb{R}_+$, along the rotation axis $\hat {\bf s}$. The kernel can then be written as
\begin{equation}  
  G_{\textrm{lat}}({\bf r}, {\bf r}';t) = g\left(d_{\perp}({\bf r}, {\bf r'};  {\hat {\bf s}}) \right).
  \label{eq:Ghoriz}
\end{equation}
and the corresponding redistribution operator is labeled as $\hat {\cal D}_\textrm{lat}$. 

Let us make a small digression here and note that, in the case of a phase delay between heating and reradiation, it is possible to incorporate the lag into the presented framework by using coordinates shifted horizontally w.r.t.~the rotation axis $\hat {\bf s}$. The kernel for that case would be written as
$$    
  G_{\textrm{shift}}({\bf r}, {\bf r}';t) = 
  G({\bf R}_{\phi {\hat {\bf s}}} {\bf r}, {\bf r'}; t)\>,
$$
where $\phi$ is the angle by which the location of irradiated element is shifted relative to the location of the radiating element, and ${\bf R}_{\pmb{\omega}}$ is the rotation matrix about the axis of rotation $\pmb{\omega}$. We expect that the angle $\phi$ is positive and proportional to the angular velocity of the star; in general it can also depend on the position and time, as long as the quasi-stationarity of redistribution assumed here is not violated. Note that the horizontal shift does not affect the overall energy balance.

We can describe individual redistribution processes by the corresponding operator $\hat {\cal D}_i$ and form the overall redistribution operator $\hat {\cal D}$ as a weighted sum of individual $\hat {\cal D}_i$:
\begin{equation}
  \hat {\cal D} = \sum_i w_i \hat {\cal D}_i,\qquad \sum_i w_i = 1 \>,
  \label{eq:weights}
\end{equation}
where $w_i$ are positive real numbers. We can view Eq.~(\ref{eq:weights}) as the decomposition of the redistribution operator $\hat {\cal D}$ into generators of a certain type of redistribution, where the weights quantify the amount of energy redistributed by the corresponding process. 

The radiosity for \citeauthor{wilson1990}'s model can be obtained by substituting $F_0'$ from Eq.~(\ref{eq:new_exitance}) into Eqs.~(\ref{eq:wilson_reflection}):
\begin{equation}
  F_{\rm out} = 
  F_0 + \left[
    \hat{\cal D}(\mathbb{1} - \hat\Pi)  +  \hat \Pi 
  \right] \hat {\cal L}_{\rm LD} F_{\rm out}\>.
  \label{eq:wilson_irrad}
\end{equation}
In turn, the updated intrinsic exitance $F_0'$ is given by the radiosity:
\begin{equation}
  F_0' = F_0  + \hat{\cal D}(\mathbb{1} -\hat\Pi) \hat{\cal L}_{\rm LD} F_{\rm out}\>.
\end{equation}
Similarly, the irradiation for the Lambertian reflection model can be written as
\begin{equation}  
  F_{\rm in} = 
  \hat {\cal L}_{\rm LD} F_0 + 
  \left[
  \hat {\cal L}_{\rm LD}\hat {\cal D} (\mathbb{1} - \hat \Pi) 
  + \hat{\cal L}_{\rm L} \hat \Pi
  \right] F_{\rm in}\>.
  \label{eq:lambert_irrad}
\end{equation}
Eqs. (\ref{eq:wilson_irrad}) and (\ref{eq:lambert_irrad}) are  the main theoretical results of the paper and represent a unification of specific reflection scheme and irradiation redistribution under one irradiation framework. The solution of these equations determines the intrinsic exitance $F_0'$ and radiosity $F_{\rm out}$:
\begin{equation}
  F_0' = F_0 + \hat{\cal D} (\mathbb{1}- \hat \Pi) F_{\rm in}\>,\qquad
  F_{\rm out} = F_0' + \hat \Pi F_{\rm in} \>.
\end{equation}
The solutions of the irradiation models, i.e., the updated exitance $F_0'$ and radiosity $F_{\rm out}$, determine the \emph{bolometric} intensity of the radiating bodies. For \citeauthor{wilson1990}'s model, the limb-darkened intensity from Eq.~(\ref{eq:intesities}) yields
\begin{equation}
    I(\hat {\bf e}, {\bf r}) = 
    \frac{F_{\rm out}({\bf r})}{D_0({\bf r})} D(\hat {\bf e}, {\bf r})
\end{equation}
while for the Lambertian reflection model we get
\begin{equation}
    I(\hat {\bf e}, {\bf r}) = 
    \frac{F'_0({\bf r})} {D_0({\bf r})} D(\hat {\bf e}, {\bf r}) + 
    \frac{1}{\pi} \left(F_{\rm out}({\bf r})-F'_0({\bf r})\right) \>.
\end{equation}

The presented Lambertian irradiation (reflection and redistribution) model is an \emph{exact} bolometric description of this process under three assumptions: (1) Lambertian reflection is wavelength independent, (2) redistribution does not affect the limb darkening of the surface, and (3) intrinsic exitance is not affected by the irradiation.

To quantify the impact of Lambertian correction, we can expand the updated intrinsic emission  $F_0'$ and radiosity $F_{\rm out}$. For \citeauthor{wilson1990}'s model, we get
\begin{align}
  F_{\rm out} = &
  F_0 
  + \left[\hat\Pi + \hat{\cal D}(\mathbb{1} - \hat\Pi)\right] \hat {\cal L}_{\rm LD} F_0 + \nonumber \\
  &+ \underline{\left[\hat\Pi + \hat{\cal D}(\mathbb{1} - \hat\Pi)\right] \hat {\cal L}_{\rm LD}\left[\hat\Pi + \hat{\cal D}(\mathbb{1} - \hat\Pi)\right] \hat {\cal L}_{\rm LD} F_0} + \ldots\>, \\
  F_0' =&
  F_0 
  + \hat{\cal D}(\mathbb{1}-\hat\Pi) \hat{\cal L}_{\rm LD} F_0 + \nonumber\\
  &+ \underline{\hat{\cal D}(\mathbb{1}-\hat\Pi) \hat{\cal L}_{\rm LD}\left[\hat\Pi + \hat{\cal D}(\mathbb{1} - \hat\Pi)\right]  \hat {\cal L}_{\rm LD} F_0} + \ldots\>,
\end{align}
whereas in the redistribution models based on the Lambertian reflection, the expansions are written as
\begin{align}
  F_{\rm out} =&
  F_0 
  + \left[\hat\Pi + \hat{\cal D}(\mathbb{1} - \hat\Pi)\right] \hat {\cal L}_{\rm LD} F_0 \nonumber\\
  &+ \underline{\left[\hat\Pi + \hat{\cal D}(\mathbb{1} - \hat\Pi)\right] \left[\hat {\cal L}_{\rm LD} \hat {\cal D} (\mathbb{1} - \hat\Pi) + \hat{\cal L}_0 \hat\Pi\right] \hat {\cal L}_{\rm LD} F_0} + \ldots\>, \\
  F_0' =& 
  F_0 
  + \hat{\cal D}(\mathbb{1} - \hat\Pi) \hat{\cal L}_{\rm LD} F_0 \nonumber\\
  & + \underline{\hat{\cal D}(\mathbb{1} - \hat\Pi) \left[\hat{\cal L}_{\rm LD} \hat {\cal D} (\mathbb{1} - \hat \Pi) + {\cal L}_0 \hat \Pi\right] \hat {\cal L}_{\rm LD} F_0} + \ldots\>.
\end{align}
By comparing the expressions for $F_{\rm out}$ ($F_0'$) in different reflection approaches we see that they differ in the underlined second-order terms. The difference is typically small and likely not measurable at the current level of precision. However, it is conceptually important as it corresponds to a different physical description of the surface boundary energy balance.

\subsection{Lossy reflection--redistribution models}

Energy is conserved when the difference between the total emitted flux $L_{\rm out}$ and the total incident flux $L_{\rm in}$ equals the total intrinsic flux $L_0$:
\begin{equation}
  L_{\rm out} - L_{\rm in}  = L_{\rm 0} \qquad
  L_* = \int_{\cal M} F_*({\bf r}) {\rm d} A ({\bf r}),
  \quad \mathrm{where~}
  * = {\rm out,in,0}\>.
\end{equation}
If we want to account for the processes that are not included in the energy balance (such as scattering), then the total flux per Eq.~(\ref{eq:redistr_op_property}) is not conserved. The losses can occur at different levels:
\renewcommand*{\descriptionlabel}[1]{\hspace{\labelsep} #1}
\begin{description}
\item[(a)] the decrease in the non-reflected part of the incident light at the surface of the irradiated star, described by the scalar function $\xi({\bf r}) \in [0,1]$:
\begin{equation}
  \delta F_0 = \hat {\cal D}(1 - \hat \Pi)\hat\Xi F_{\rm in}
\end{equation}
where we introduce the auxiliary operator $\hat \Xi:f\mapsto \xi f$ to describe the losses;
\item [(b)] the decrease of energy in the interior of the irradiated star, described by a ``lossy'' redistribution operator $\hat {\cal D}'$:
\begin{equation}
\delta F_0 = \hat {\cal D}'(\mathbb{1} - \hat \Pi)F_{\rm in}\>,
\end{equation}
where the redistribution operator has the following property for an arbitrary positive function $F$:
\begin{equation}
\int_{\cal M} [\hat {\cal D}' F]({\bf r}) {\rm d} A({\bf r}) \le 
\int_{\cal M} F({\bf r}) {\rm d} A({\bf r})\>; \quad \mathrm{or}
\end{equation}

\item [(c)] the decrease in the emergent light at the surface of the radiating star, described analogously to case (a):
\begin{equation}
  \delta F_0 = \hat \Xi \hat {\cal D}(\mathbb{1} - \hat \Pi)F_{\rm in} \>.
\end{equation}
which is just a specific case of the previous with $\hat {\cal D}' = \hat \Xi \hat {\cal D}$, from the modeling point of view.

\end{description}

Case (a) could be used to describe losses due to scattering from the surface that are not taken into account by the reflection; case (b) could mimic the absorption of energy by processes inside of the star that violate flux conservation in a semi-stationary regime, e.g., altering the internal dynamics of the envelope in convective stars \citep{rucinski1969}; and case (c) could model obstructions for the re-emission of redistributed irradiation (e.g., some irregularities on the surface of the stars in the form of spots, convective cells, etc.). How to translate the mentioned processes into the redistribution model is beyond the scope of this paper.

If the loss and reflection coefficients $\xi$ and $\rho$ are constant over the surface, we can express flux losses $L_{\rm loss} = L_0 - (L_{\rm out} - L_{\rm in})$ in the cases (a) and (b) as
$$
  L_{\rm loss} = L_{\rm in} (1 - \rho) (1 - \xi)\>, 
$$
where $L_{\rm in}$ cannot be written in a simple form as it depends on the radiosity operators. Setting $\xi=1$ eliminates the losses. Flux conservation can be described by the three fractions, all with respect to the incident flux: the part of the incident flux reflected from the surface, $R_\mathrm{refl}$; the part of the flux absorbed and then redistributed across the surface, $R_\mathrm{redistr}$; and the part of the flux that is lost at the surface, $R_{\rm lost}$:
\begin{equation}
  R_{\rm refl} + R_{\rm redistr} + R_{\rm lost} = 1 \>,
\end{equation}
where $R_{\rm refl} \equiv \rho$, $R_{\rm redistr} \equiv \xi (1-\rho)$, and $R_{\rm lost} \equiv (1-\xi) (1-\rho)$.

\subsection{Effective temperatures and non-bolometric observations}\label{ch:temp_nonbol}

The local effective temperature of a surface element is defined by the radiosity $F_{\rm out}$ of blackbody emission:
\begin{equation}
  T_{\rm eff}({\bf r}) = \frac{1}{\sigma} F_{\rm out}({\bf r})^\frac{1}{4} \>,
\end{equation}
where $\sigma$ is the Stefan--Boltzmann constant. In the absence of reflection, the radiosity of a star equals the intrinsic exitance $F_0$ and the corresponding intrinsic effective temperature is denoted by $T_{\rm eff,0}$, which is distributed across the surface according to the adopted gravity-darkening model (i.e.~\citet{zeipel1924} for radiative photospheres). In the presence of reflection and redistribution, the intrinsic exitance gets updated to $F_0'$ and the radiosity equals $F_{\rm out}$. The effective temperature associated with the updated intrinsic exitance is
\begin{equation}
    T_{\rm eff,0}'({\bf r}) =
    T_{\rm eff,0} ({\bf r})
    \left(\frac{F_0'({\bf r})} {F_0({\bf r})} \right)^\frac{1}{4}
\end{equation}
and the effective local temperature  associated with radiosity is equal to
\begin{equation}
  T_{\rm eff} ({\bf r}) 
  = T_{\rm eff,0}' ({\bf r}) 
  \left (\frac{F_{\rm out} ({\bf r})} {F_0'({\bf r})} \right)^\frac{1}{4}
  = T_{\rm eff,0}' ({\bf r}) 
  \left (1 + \rho({\bf r})\frac{F_{\rm in} ({\bf r})} {F_0'({\bf r})} \right)^\frac{1}{4} \>.
  \label{eq:Teff}
\end{equation}

The results of the irradiation framework presented here are bolometric quantities, which are wavelength-independent. Because of that, we can only synthesize wavelength independent (bolometric) observations. That said, the treatment described above lends itself readily to the wavelength-dependent re-emission approximation analogous to that of \citet{wilson1990}. We outline this procedure for a given wavelength-dependent plane-parallel stellar atmospheric model, distribution of local effective temperature $T_{\rm eff,0}$ and other properties of the atmosphere across the isolated star:
\begin{enumerate}
\item The atmospheric model determines the spectral intensity on the surface of the star given by
\begin{equation}
 I_{\rm atm}(\lambda, \cos\theta, T, \ldots) \>,
\end{equation}
where $\lambda$ is the wavelength, $\theta$ is the angle from the normal to the plane, $T$ is the local effective temperature, and "$\ldots$" marks all other parameters that determine properties of the atmosphere. From the spectral intensity we calculate intrinsic exitance $F_0$ at each point ${\bf r}$ of the surface,
\begin{equation}
    F_0({\bf r}) = 
   2\pi \int_0^\infty {\rm d} \lambda \int_0^1 {\rm d} \mu\, \mu 
    I_{\rm atm}(\lambda, \mu, T_{\rm eff,0}({\bf r}), \ldots)\>.
\end{equation}
\item From the intrinsic exitance $F_0$, albedo $\rho$, and parameters determining redistribution, we calculate using here the presented irradiation framework using the updated exitance $F_0'$ and radiosity $F_{\rm out}$.

\item The radiosity $F_{\rm out}$ determines the local effective temperature $T_{\rm eff}$ (Eq. \ref{eq:Teff}). Following \citet{wilson1990}, we may use the effective temperature as the new local temperature in the spectral intensity,
\begin{equation}
    I_{\rm atm}(\lambda, \cos\theta, T_{\rm eff}({\bf r}), \ldots) \>,
\end{equation}
in order to calculate non-bolometric observables.
\end{enumerate}

The outlined procedure is effectively a reinterpretation of bolometric results in the spectral sense and can only be seen as a rough approximation for the truly wavelength-dependent irradiation framework that would involve ray-tracing the light coming from each surface element to the observer, a complicated and computationally extensive scheme beyond the scope of this paper. The approximate procedure outlined above is currently the standard way of dealing with this technical issue. In order to be consistent throughout the paper, we focus purely on bolometric processes and bolometric observations, but using the described procedure one can also model passband-dependent observations.

\section{Discretization of the irradiation framework}\label{sec:discret}

In order to use the presented irradiation framework in practice, all introduced operators, i.e., radiosity $\hat {\cal L}_*$ ($*={\rm L}, {\rm LD}$), reflection $\hat \Pi$ and redistribution ${\hat D}$, and all functions defined on the surface that describe physical properties of the body, such as radiosity, intensity, and emittance, need to be discretized.

\subsection{Basic concepts behind discretization}

We start the discretization by partitioning the surface $\cal M$ into distinct subsets:
\begin{equation}
  {\cal M} = \bigcup_i {\cal S}_i\qquad 
  {\cal S}_i \cap {\cal S}_j = 0\qquad i\neq j \>.
\end{equation}
with their area equal to
\begin{equation}
    A_i = \int_{{\cal S}_i} {\rm d} A({\bf r})\>.
\end{equation}

We approximate functions defined on the surface as functions constant over any surface element ${\cal S}_i$, called the piecewise constant function over $\cal M$. A piecewise constant approximation $\tilde f$ of an integrable function $f$ defined on the surface $\cal M$ is given by
\begin{equation}
  \tilde f({\bf r}) = \sum_i f_i \chi_i ({\bf r})  \qquad
  \chi_i ({\bf r}) = \left \{
  \begin{array}{lll}
  1 &:& {\bf r} \in {\cal S}_i \\
  0 &:& \textrm{otherwise}
  \end{array}
  \right. \>,
  \label{eq:fun_discr}
\end{equation}
with the expansion coefficients $f_i$ expressed as
\begin{equation}
    f_i = 
    \frac{1}{A_i} \int_{{\cal S}_i} f({\bf r}) {\rm d} A({\bf r}) \>,
\end{equation}
where $\chi_i({\bf r})$ is a characteristic function of ${\cal S}_i$ on $\cal M$. The set $\{\chi_i ({\bf r}) \}$ is a functional basis of piecewise constant functions. We treat the vector ${\bf f} = (f_i)$ as discretized version of the function $f$.

Here, the considered operators are linear and therefore can be written in the form
\begin{equation}
     \hat {\cal O} f({\bf r}) = 
  \int_{\cal M} H({\bf r}, {\bf r}') f({\bf r}') {\rm d} A({\bf r'}) \>,
  \label{eq:exact_map}
\end{equation}
where $H({\bf r}, {\bf r}')$ is a kernel function that depends on the operator we are considering. We have a piecewise constant function $f=\sum_i f_i \chi_i$ with expansion coefficients $f_i$. We approximate its image $\hat {\cal O} f$ by a piecewise constant function $\sum_i f_i' \chi_i$ with the expansion coefficients given by
\begin{equation}
  f_i' = \sum_j O_{i,j} f_j\>, 
  \label{eq:map_discr}
\end{equation}
where the matrix elements $O_{i,j}$ are expressed as
\begin{equation}
  O_{i,j}= \frac{1}{A_i}
    \int_{{\cal S}_i}  
    \int_{{\cal S}_j} H({\bf r}, {\bf r}') {\rm d} A({\bf r'})\, {\rm d} A({\bf r}) \>.
    \label{eq:mat_elem}
\end{equation}
The matrix ${\bf O} = [O_{i,j}]$ is the discretized version of the operator $\hat {\cal O}$. In the case of the radiosity operator, $O_{i,j}$ are the generalizations of the view factors \citep{modest2003}. For the discretized version of the redistribution operator $\hat{\cal D}$, denoted by the matrix ${\bf D}= [D_{i,j}]$, the flux conservation takes the form
\begin{equation}
  \sum_i A_i D_{i,j} = A_j \>.
\end{equation}

\subsection{Calculations on the triangular surfaces and practical considerations}

The irradiation framework is implemented by extending the open-source package \phoebe, where the working surface $\cal M$ is a mesh of triangles that approximates the true shape of astrophysical bodies. It supports different geometrical bodies, e.g., aligned and misaligned Roche0shaped stars and isolated rotating stars. Triangular discretization of Roche-shaped bodies was already used in, e.g.  \cite{hendry2000} and \cite{pribulla2012}, just to name a few. We consider two discretization schemes for operators, per-triangle discretization and per-vertex discretization, as described in \citet{prsa2016}.  In both cases, we simplify the expressions for matrix elements Eq. (\ref{eq:mat_elem}) to
\begin{equation}
 O_{i,j} \approx A_j H({\bf r}_i, {\bf r}_j) \>,
\end{equation}
where ${\bf r}_i$ and ${\bf r}_j$ are the surface element locations and $A_i$ are the corresponding areas.

\begin{figure}[!htb]
\vbox{%
\centering
\begin{minipage}[l]{5cm}
\centering
\includegraphics[width=5cm]{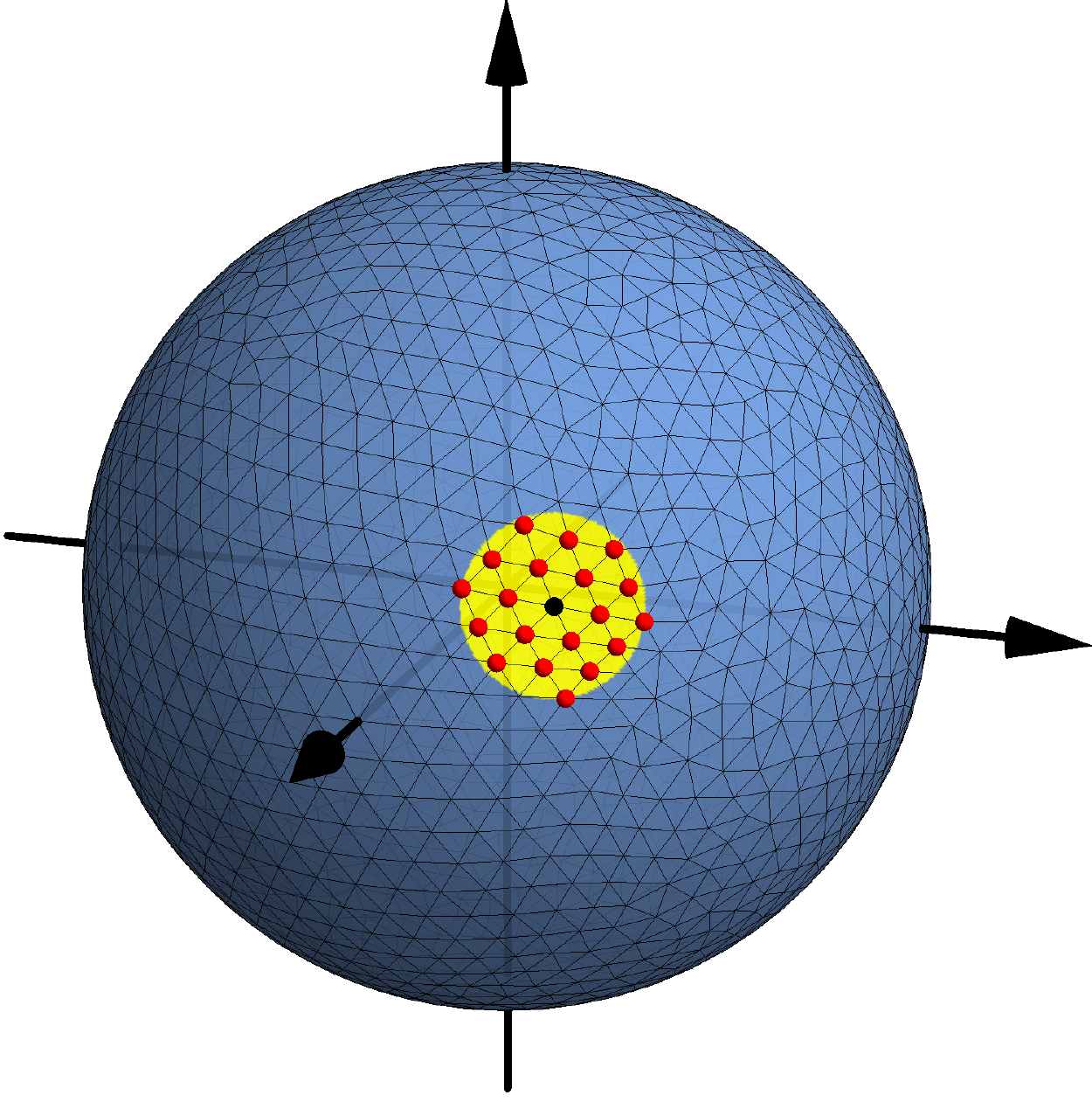}\\
(a)
\end{minipage}
\hspace{2cm}
\begin{minipage}[r]{5cm}
\centering
\includegraphics[width=5cm]{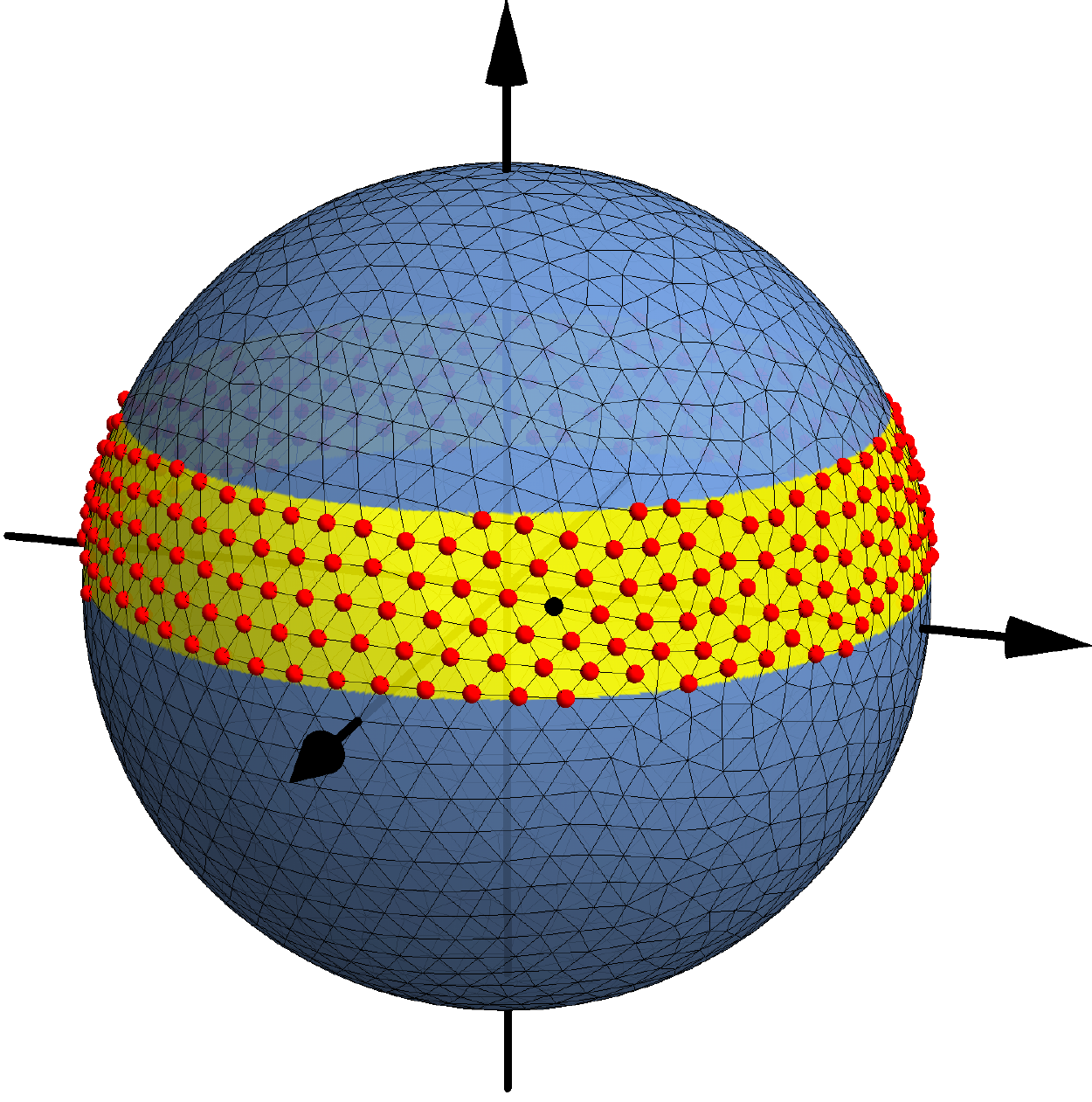}\\
(b)
\end{minipage}
}

\caption{Schematic figure of the local (a) and latitudinal (b) redistribution of the flux incident on the surface element surrounding the vertex depicted in black and emitted from the surface elements that surround the vertices depicted in red.}
\label{fig:redistr}
\end{figure}

In the decomposition of the redistribution operators, we need the distances across the surfaces. The calculation of distances on the triangular meshes is computationally very expensive; see, e.g., \citet{martinez2005}. For astrophysical bodies close to spherical, we can frequently approximate the distances by those on the sphere, which is computationally tractable. Well-detached stars in a binary configuration certainly fall into this category. Consider our object of interest packed inside a sphere of radius $R$ and center ${\bf c}$, satisfying
\begin{equation}
  \sum_i \left(\|{\bf r}_i - {\bf c} \|^2 - R^2\right)^2 = {\rm min.}
\end{equation}
Next, define an operator to obtain the radial unit vector at a point on the sphere w.r.t.~the center ${\bf c}$,
\begin{equation}
  \hat{\mathbb{P}} {\bf r} =\frac {{\bf r} - {\bf c}}{\| {\bf r} - {\bf c} \|} \>.
\end{equation}
The geodesic distance between the two points on the mesh, $({\bf r}_1, {\bf r}_2)$, is approximated by the distance between the corresponding points on the sphere:
\begin{equation}
  d({\bf r}_1, {\bf r}_2) = 
  R \arccos(\,(\hat{\mathbb{P}} {\bf r}_1) \cdot (\hat{\mathbb{P}} {\bf r}_2)\,) \>.
\end{equation}
Furthermore, if the local curvature of the mesh is small, i.e., the mesh points are dense and the distances between neighboring points considered for local redistribution do not differ much from the geodesic, we can approximate the distances with the Euclidean form: 
\begin{equation}
  d({\bf r}_1, {\bf r}_2) \approx || {\bf r}_1 - {\bf r}_2 || \>.
\end{equation}
The deviation from the surface, measured along the axis $\boldsymbol\phi$, ($\|\boldsymbol{\phi}\|=1$), is then
\begin{equation}
  d_\perp ({\bf r}_1, {\bf r}_2) =
  R \arccos\left(\sqrt{(1-u_1^2)(1-u_2^2)} + u_1 u_2\right),
  \qquad
  u_i  =  {\boldsymbol\phi} \cdot \hat{\mathbb{P}} {\bf r}_i\>. 
\end{equation}
The local and latitudinal redistribution models, based on $d$ and $d_\perp$, are schematically depicted in Fig.~\ref{fig:redistr}. The flux incident on the surface element centered on the vertex depicted in black is redistributed over the surface elements depicted in yellow, associated with the vertices depicted in red. The fraction of the incident flux that is redistributed over the elements depends on the weight function discussed before.

\subsection{Irradiation parameters}\label{sec:irrad_par}

For most practical cases, it suffices to assume discrete redistribution models (i.e.~local, latitudinal and global) with constant reflection coefficients.  The irradiation parameters $\rho_{\rm refl}$, $\rho_{\rm loc}$, $\rho_{\rm lat}$, and $\rho_{\rm uni}$, which are associated with the reflection and the local, latitudinal, and uniform redistributions, respectively, are used to construct operator weights (cf.~Eq.~\ref{eq:weights}):
\begin{equation}
  w_{\rm loc} = \frac{\rho_{\rm loc}}{1-\rho_{\rm refl}}\>,\quad
  w_{\rm lat} = \frac{\rho_{\rm lat}}{1-\rho_{\rm refl}}\>,\quad
  w_{\rm uni} = \frac{\rho_{\rm uni}}{1-\rho_{\rm refl}} \>,
\end{equation}
which are non-negative and sum up to $1$. In turn, they determine the redistribution matrix for a given object,
\begin{equation}
  {\bf D} = 
  w_{\rm loc} {\bf D}_{\rm loc} + 
  w_{\rm lat} {\bf D}_{\rm lat} + 
  w_{\rm uni} {\bf D}_{\rm uni}  \>.
\end{equation}
The principal advantage of these irradiation parameters, i.e., $\rho_{\rm refl}$, $\rho_{\rm loc}$, $\rho_{\rm lat}$ and $\rho_{\rm uni}$, is that they add up to 1 for each body separately (assuming no losses), and each individual parameter represents the fraction of the total incoming flux redistributed by the given irradiation process.

Notice that the uniform redistribution matrix for an $i$th body ${\bf D}_{{\rm uni}, i}$ can be expressed as a projection:
\begin{equation}
  {\bf D}_{{\rm uni}, i} = \frac{1}{A_i} [1,\ldots, 1]^T [A_{i,1}, A_{i,2}, \ldots, A_{i,N_i} ]  \>,
\end{equation}
where $A_{i,j}$ is the area of the $j$th surface element $(j\in [1,N_i])$ on the $i$th body and $A_i=\sum_j A_{i,j}$ is the total area of the body. Using this property, we can significantly speed up calculations related to uniform redistribution. This is taken into account in the implementation of the irradiation framework in the extension of \phoebe 2.1.

\subsection{Solving discrete reflection--redistribution equations}

We follow the presented discretization procedure and approximate all operators by matrices and all functions defined on the surface by vectors with their entries representing average quantities on surface elements. The discretized limb-darkened $\hat {\cal L}_{\rm LD}$ and Lambertian $\hat {\cal L}_{\rm L}$ radiosity operators are represented by the matrices ${\bf L}_{\rm LD}$ and ${\bf L}_{\rm LD}$, respectively, and the redistribution operator $\hat {\cal D}$ is described by the matrix ${\bf D}$ and the reflection operator $\hat\Pi$ is approximated by $\mathbf{\Pi}$. The radiosity $F_{\rm out}$, intrinsic exitance $F_0$, updated exitance $F_0'$, and irradiation $F_{\rm in}$ are approximated by the vectors ${\bf F}_{\rm out}$, ${\bf F}_{\rm 0}$, ${\bf F}_{\rm 0}'$ and ${\bf F}_{\rm in}$, respectively.

The essential reflection--redistribution equations are given by Eq.~(\ref{eq:wilson_irrad}) for \citeauthor{wilson1990}'s reflection and by Eq.~(\ref{eq:lambert_irrad}) for the Lambertian reflection model. Following the discretization rules, these can be written in matrix form as
\begin{align}
  {\bf F}_{\rm out} = & {\bf G}_{\rm W} + {\bf Q}_{\rm W}  {\bf F}_{\rm out}\> \quad & \textrm{for \citeauthor{wilson1990}'s model,}
  \label{eq:mat_eq_wilson}
  \\
  {\bf F}_{\rm in} = & {\bf G}_{\rm L} + {\bf Q}_{\rm L}  {\bf F}_{\rm in} \quad & \textrm{for Lambertian model,}
  \label{eq:mat_eq_lambert}
\end{align}
where we, for compactness, introduce the auxiliary vectors ${\bf G}_{\rm W}$ and ${\bf G}_{\rm L}$ given by
\begin{equation}
    {\bf G}_{\rm L} = {\bf F}_0\qquad
    {\bf G}_{\rm W} = {\bf L}_{\rm LD} {\bf F}_0 \>,
\end{equation}
and the matrices ${\bf Q}_s$ written as
\begin{equation}
{\bf Q}_{\rm W} = \left[{\bf D}(\mathbb{1}- \mathbf{\Pi}) + \mathbf{\Pi}\right] {\bf L}_{\rm LD}\qquad
{\bf Q}_{\rm L} = {\bf L}_{\rm LD}{\bf D}(\mathbb{1}- \mathbf{\Pi}) + {\bf L}_{\rm L}\mathbf{\Pi}\>.
\label{eq:matQ}
\end{equation}
The vectors ${\bf G}_{\rm W}$ and ${\bf G}_{\rm L}$ represent the intrinsic exitance and limb-darkened irradiated exitance, respectively, whereas the matrices ${\bf Q}_s$ represent the products of the discretized radiosity operator, reflection coefficients, and the redistribution operator associated with a given reflection--redistribution scheme described by equations (\ref{eq:wilson_irrad}) and (\ref{eq:lambert_irrad}).

By design, the matrices ${\bf Q}_s$ scale linearly with the discretized radiosity operators and, consequently, its norm is $\leq 1$. Because of that, Eqs.~(\ref{eq:mat_eq_wilson})--(\ref{eq:mat_eq_lambert}) are convergent and can be solved iteratively:
\begin{equation}
    {\bf F}^{(k+1)} = {\bf G}_s + {\bf Q}_s  {\bf F}^{(k)} 
    \qquad \textrm{for}~k=0,1,\ldots\>,
    \label{eq:irrad_iter}
\end{equation}
with $s={\rm W}, {\rm L}$ labeling the reflection--redistribution scheme and using the initial condition ${\bf F}^{(0)} = {\bf G}_s$. The vector ${\bf F}$ represents ${\bf F}_{\rm out}$ and ${\bf F}_{\rm in}$ in \citeauthor{wilson1990}'s and the Lambertian reflection model, respectively. Accurate irradiation calculations can be very time consuming. It involves constructing all matrices and solving a large sparse system of linear equations, as described in this section. Therefore, it is useful to have an approximate model to determine the magnitude of irradiation effects in order to decide whether they need to be taken into account given the required precision. We provide a detailed discussion in Appendix \ref{sec:1d_model}.

\subsection{Time complexity of irradiation framework}

We estimate the time complexity of the different phases of irradiation framework as implemented in the extension of \phoebe, i.e., triangulating of bodies, obtaining properties of the mesh (area of elements, total area, total volume, ...), calculating radiosity operator matrices describing reflection, calculating redistribution operator matrices, and finally, solving the linear system. 

Let us assume that we have $m$ convex bodies and the surface of the $i$th body is partitioned into $N_i$ elements, which in our case are triangles. The generation of triangular mesh covering the surface of bodies and the calculation of the properties of the mesh are a standard part of \phoebe 2.1 and are performed in $\sum_i {\cal O}(N_i)$ operations, where ${\cal O}$ signifies the limiting behavior of a function; see, e.g., in \cite{sirca2018}. The radiosity operator matrices ${\bf L}_{\rm LD}$ and ${\bf L}_0$ are sparse matrices in all practical cases, because the visibility between surface elements is frequently obstructed. Their construction in general takes ${\cal O}((\sum_i N_i)^2)$ operations, but for convex bodies, the number of operations reduces to ${\cal O}(\sum_{i>j} N_i N_j)$. The redistribution matrix ${\bf D}$ has a block diagonal form, where each block corresponds to a separate body. A block associated with the $i$th body has a dimension $N_i \times N_i$. In order to calculate all of the blocks, we need $\sum_i {\cal O}(N_i^2)$ operations.  By setting $N_i=N$, the computational costs of constructing the redistribution and radiosity operator matrices are equal to ${\cal O}(m N^2)$  and ${\cal O}(m(m-1)N^2)$, respectively. Notice that the latter grows quadratically with the number of bodies, whereas the former grows only linearly. The matrices ${\bf Q}_s$  ($s={\rm W}, {\rm L}$) (Eq.~\ref{eq:matQ}) are of dimension $M\times M$, where $M=\sum_i N_i$ is the number of surface elements, and are typically still sparse with $N_\textrm{nonzero} = {\cal O}(\sum_{i>j} N_i N_j) + \sum_i {\cal O}(N_i^2)$ non-zero elements. The system of equations determined by the matrices ${\bf Q}_s$ are solved iteratively. Assuming we need $N_{\rm it}$ iterations, the solution can be found in $N_{\rm it} N_\textrm{nonzero}$ operations.

\begin{figure}[!htb]
\centering
\includegraphics[width=12cm]{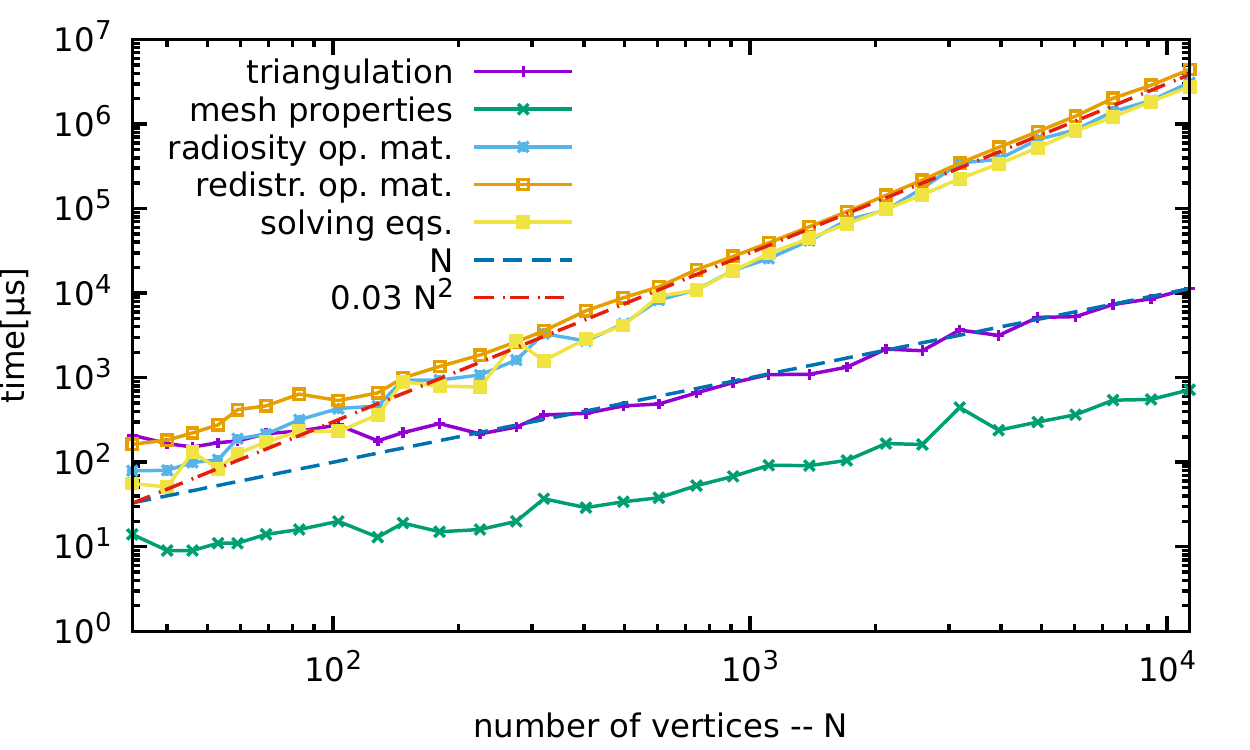}
\caption{Time needed for different phases of the irradiation framework in a binary system as a function of the number of triangles $N$ on a computer with an Intel i7-4600U CPU processor at 2.10GHz using a single core. The binary system is composed of two identical Roche-shaped stars determined by the following parameters: mass ratio $q = 1$, synchronicity parameter $F_{\rm sync} = 1$, star separation $\delta= 1$, and equivalent radius of the stars $r_{\rm equiv}\doteq 0.162818$. We are discussing redistribution without losses using the linear weight function with the threshold value $l/R=0.2$ and equally weighted local, latitudinal, and uniform redistributions: $w_{\rm loc}=w_{\rm lat}=w_{\rm uni}=1/3$. The reflection is performed with an albedo of both stars $\rho_{\rm refl}=0.3$ and with the linear limb darkening at the coefficient $x=0.3$.}
\label{fig:redistr_timing}
\end{figure}

The times needed for the different phases of the irradiation framework as a function of the number of surface elements $N=N_i$ for a simple binary system of two identical Roche-shaped stars are depicted in the  Fig.~\ref{fig:redistr_timing}. Notice that the times needed to generate individual matrices and obtain a solution are of the same order of magnitude and are by far the most costly part of the irradiation framework. These times have a clear quadratic dependence on $N$ in comparison to times to generate the mesh and calculate its properties, which scale linearly with $N$.

\section{Demonstration of principles}

In this section, we demonstrate the irradiation models and underlying redistribution processes for several toy models to get a qualitative understanding of the presented irradiation models. To this end, we use \phoebe \citep{prsa2016a}, an open-source package for modeling eclipsing binaries, where we implement an irradiation framework using discretized operator as presented in Section \ref{sec:discret}: the redistribution process is modeled as a linear superposition of local, latitudinal, and global redistribution; the irradiation parameters, i.e., $\rho_{\rm refl}$, $\rho_{\rm loc}$, $\rho_{\rm lat}$, and $\rho_{\rm uni}$, are constants chosen for each body separately. If the irradiation parameters do not add up to 1, we have irradiation losses.

\subsection{Irradiance and radiosity in a two-sphere system}

\begin{figure}[!htb]
\centering
\begin{tabular}{cc}
  \rotatebox[origin=c]{90}{uniform}&
  \begin{minipage}[c]{14cm}
  \includegraphics[width=14cm]{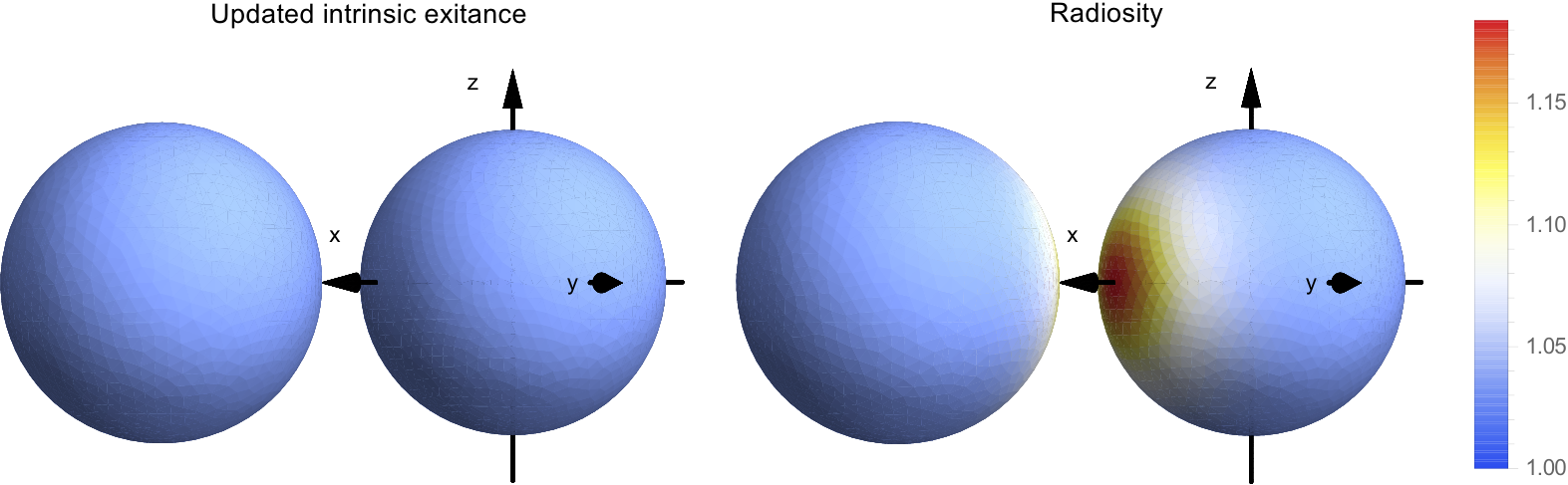}
  \end{minipage}\\
  \rotatebox[origin=c]{90}{local}&
  \begin{minipage}[c]{14cm}
  \includegraphics[width=14cm]{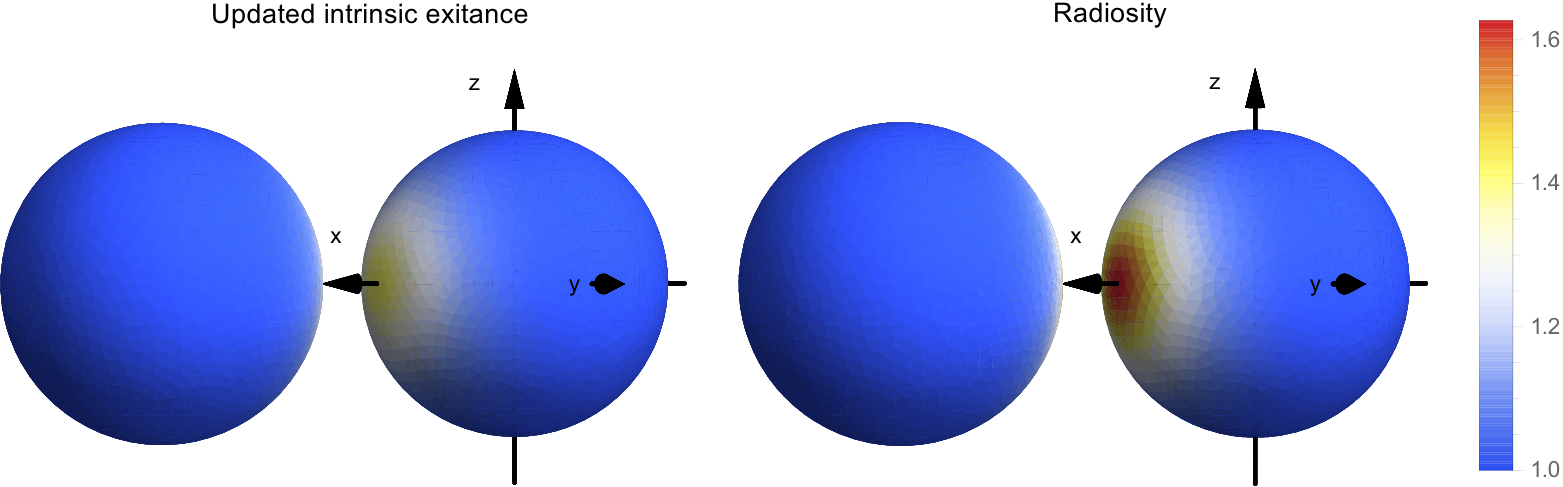}
  \end{minipage}\\
  \rotatebox[origin=c]{90}{latitudinal}&
  \begin{minipage}[c]{14cm}
  \includegraphics[width=14cm]{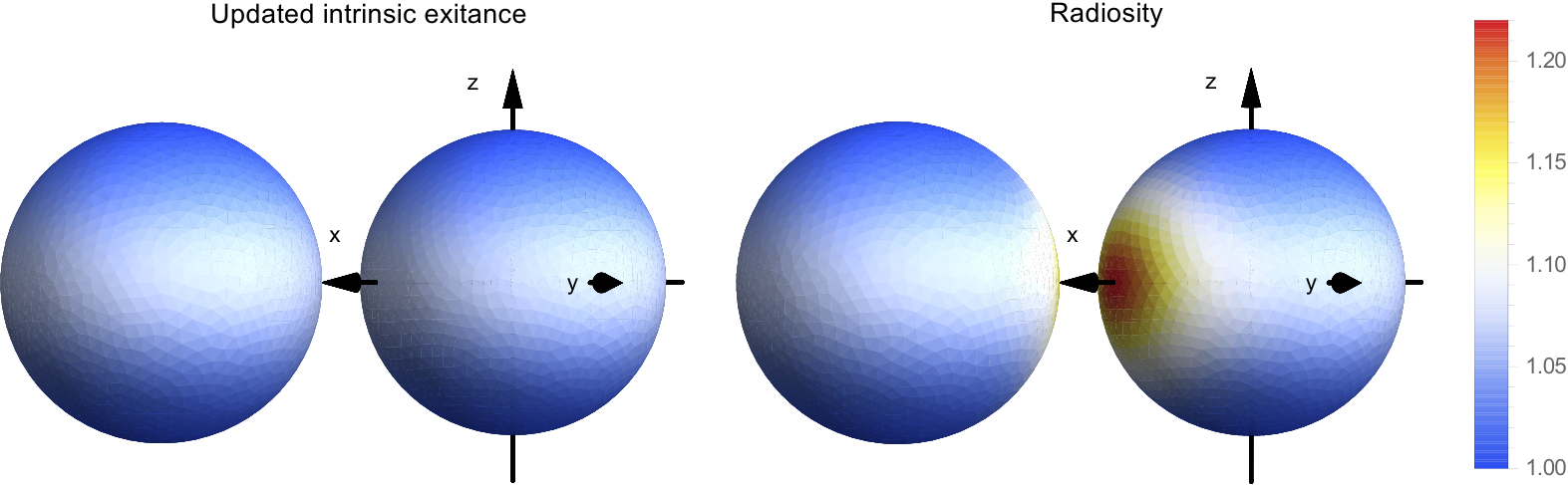}
  \end{minipage}
\end{tabular}
\caption{The changes in updated intrinsic exitance and radiosity as a function of redistribution model. Each row corresponds to a specific redistribution type for a system of two spherical stars. Both stars have a relative size $R=1$ and the centers of the stars are separated by $L=2.5$. The Lambertian reflection approach has been implemented with reflection and linear limb darkening coefficients of $\rho_{\rm refl}=0.3$ and $x=0.3$, respectively. The intrinsic emission for both stars is set to $F_0=1$ and we use a linear weight function $g$ with the threshold value of $l/R = 0.2$.}
\label{fig:redistr_demo}
\end{figure}

Consider a system of two identical spheres with a constant exitance $F_0$. The consequence of mutual irradiation is the updated exitance $F_0'$ and radiosity $F$ of the two spheres. We compute it by using Lambertian reflection and a specific redistribution model with a linear weight function $g(x) = 1 - x/l$ without considering losses. The results are depicted in Fig.~\ref{fig:redistr_demo} as a density plot of $F_0'$ and $F$ across the surfaces of the spheres. Common to all redistribution models is that the increase of the reflection coefficient decreases the incident flux redistributed over the surface and, in consequence, a decrease in $F_0'$; and the increase of the area over which the incoming flux is redistributed makes $F_0'$ more uniform across the surface and, on average, decrease in size. 

In order to highlight the effects of both processes, reflection and redistribution, we choose a reflection coefficient $\rho=0.3$, which is large enough to have notable reflection and small enough to enable significant redistribution. Fig.~\ref{fig:redistr_demo} shows that the uniform redistribution produces a uniform $F_0'$ and represents a bias value for radiosity, as indicated by Eq.~(\ref{eq:new_exitance}); the local redistribution generally gives the largest $F_0'$ and $F$, and both distributions have a similar shape; and lastly, the latitudinal redistribution increases exitance at latitudes that are most strongly illuminated. 

A mean-field approximation of two-sphere case here discussed with a focus on average radiosity and irradiance is presented in Appendix \ref{sec:1d_model} and can be used to check the order of magnitude of the reflection--redistribution effects.

\subsection{Detecting and discriminating irradiation effects} \label{sec:detect_discriminate}

We apply the introduced irradiation models by calculating LCs for a binary star with redistribution switched both on and off. We consider simplified reflection in which reflection coefficients are constant across a body. For the presentation purposes, the bodies are spherical, but the theory is valid for any geometrically defined surface. In addition, we discuss how well we can discriminate between the different redistribution effects based on the LCs.

Consider a model LC with certain redistribution parameters ${\bf x}  \in \mathbb{R}^p$ calculated at $N$ time stamps: ${\bf C}({\bf x}) = [C_i({\bf x})]_{i=1}^N  \in \mathbb{R}^N$. In the vicinity of the parameters ${\bf x}$ (i.e., for perturbed parameters ${\bf x}'= {\bf x} + \delta {\bf x} \in \mathbb{R}^p$) we can approximate this LC vector by its Taylor expansion:
$$
  {\bf C}({\bf x}') 
  = {\bf C}({\bf x}) + {\bf C}'({\bf x}) \delta {\bf x} + 
  \mathcal{O}(\|\delta {\bf x}\|^2)\>.
$$
The vector norms in use here are denoted by $\| \cdot \|$ and for a vector ${\bf x}$ is defined as $\|{\bf x}\| = \sqrt{{\bf x}^T {\bf x}}$. The discrepancy between the LC vector at the parameters $\bf x$ and at the perturbed parameters ${\bf x}'$ is measured by the norm of the difference of the LC vectors, written as
\begin{equation}
  \|{\bf C}({\bf x}') - {\bf C} ({\bf x})\|^2 = 
  \delta {\bf x}^T {\bf C}'^T({\bf x}) {\bf C}'({\bf x}) \delta {\bf x} 
  + \mathcal{O}(\|\delta {\bf x}\|^3) \>.
  \label{eq:LC_discr}
\end{equation}
The discrepancy can be quantified by performing a singular value decomposition (SVD) of the LC vector derivative,  
\begin{equation}
  {\bf C}'({\bf x}) = 
  {\bf U}({\bf x}) \boldsymbol{\Sigma}({\bf x}) {\bf V}^T({\bf x})\>,
   \label{eq:LC_SVD} 
\end{equation}
where ${\bf U}({\bf x}) \in \mathbb{R}^{N\times N}$ and ${\bf V}({\bf x})\in \mathbb{R}^{p\times p}$ are orthogonal matrices and $\boldsymbol{\Sigma}({\bf x}) \in \mathbb{R}^{N\times p}$ is a diagonal matrix with singular values on the diagonal: the maximum and the minimum value on the diagonal are $\sigma_{\rm max}$ and $\sigma_{\rm min}$, respectively. For details on SVD, see, e.g.,  \cite{sirca2018}. The discrepancy of the LC vector in the limit of small perturbations is then bound by the singular values
\begin{equation}
  \sigma_{\rm min}({\bf x}) \|\delta {\bf x}\|
  \le
  \|{\bf C}({\bf x} + \delta {\bf x}) - {\bf C} ({\bf x})\|
  \le 
  \sigma_{\rm max}({\bf x}) \|\delta {\bf x}\| \>.
  \label{eq:var_C}
\end{equation}
When $\sigma_{\rm min}$ is zero, there is a linear combination of irradiation effects that do not produce changes in the light curve, resulting in a degenerate case. Note that $\sigma_{\rm min}$ and $\sigma_{\rm max}$ have a dimension and, consequently, they scale linearly with the amplitude of the light curve. 

Typically, the difference in LCs ${\bf C}({\bf x} + \delta {\bf x}) - {\bf C} ({\bf x})$ can be measured up to a certain noise level. Let us denote the discrepancy between the measured and computed LCs as $\bf N$ and treat it as noise\footnote{If the noise is uncorrelated with the standard deviation $\sigma_{\rm noise}$, the statistical averages of the vector norm of the noise and the corresponding square are $\langle \|{\bf N}\| \rangle = \sqrt{\frac{2N}{\pi}} \sigma_{\rm noise}$ and $\langle \|{\bf N}\|^2 \rangle = N \sigma_{\rm noise}^2$, respectively.}. The changes in the irradiation parameters $\delta {\bf x}$ are detectable if
\begin{equation}  
  \|\delta {\bf x}\| \ge \frac{\|{\bf N}\|}{ \sigma_{\rm max}({\bf x})} \equiv 
  \epsilon_{\rm sufficient} 
\end{equation}
and all changes in the irradiation parameters are measurable if
\begin{equation}  
  \|\delta {\bf x}\| \ge \frac{\|{\bf N}\|}{ \sigma_{\rm min}({\bf x})} \equiv 
  \epsilon_{\rm total} \>.
  \label{eq:irrad_low_limit} 
\end{equation}
To quantify how well we can discriminate between effects, we need to consider how strongly the LC varies due to changes in irradiation parameters about ${\bf x}$. We quantify the variation by the ratio between the largest and the smallest responses in the LC vector equal to the condition number \citep{sirca2018}:
\begin{equation}
  \kappa({\bf x}) = \frac{\epsilon_{\rm total}}{\epsilon_{\rm sufficient}} =
  \frac
    { \sigma_{\rm max}({\bf x})}
    { \sigma_{\rm min}({\bf x})} \ge 1 \>.
  \label{eq:cond_nr}
\end{equation}
In order to clearly separate the effect of different parameters on the LC, $\kappa$ needs to be as large as possible and $\|\delta {\bf x}\| \ge \epsilon_{\rm total}$. Note that, in the degenerate case, the conditional number $\kappa$ is infinite and so is $\epsilon_{\rm total}$. 

\subsection{Bolometric LCs for a toy binary system}

We are studying irradiation effects in a toy binary system motivated by NN Serpentis, an eclipsing binary system composed of a white dwarf (primary star - P) and red dwarf (secondary star - S) with an orbital period of 0.13 days \citep{qian2009, parsons2010}, although without its recently discovered circumbinary disk \citep{hardy2016}. The parameters of the toy system are listed in Table \ref{tab:NNserpensis}, where only the red dwarf is subjected to irradiation redistribution.  The lobes of the stars in the toy system are described in Roche geometry. The redistribution of irradiation is applied to the system as described in Sec.~\ref{sec:discret}, with irradiation in this two-body system described by 8 parameters: $\rho_{\rm refl,b}$, $\rho_{\rm loc,b}$, $\rho_{\rm lat,b}$, and $\rho_{\rm uni,b}$ for bodies $b={\rm S,P}$. 

We choose that the white dwarf's albedo to be $\rho_{\rm refl,P}=1$ and the red dwarf's albedo to be $\rho_{\rm refl,S}=0.6$. This means for the former that $\rho_{\rm loc,P}=\rho_{\rm lat,P}=\rho_{\rm uni,P} = 0$ and for the latter that it can be subjected to irradiation redistribution effects. The synthetic bolometric LCs and radial velocity curves are calculated in the limiting cases of redistribution in the red dwarf: in the absence of redistribution ($\rho_{\rm loc,S}=\rho_{\rm lat,S}=\rho_{\rm uni,S}=0$), where we are confronted with losses; in the presence of only local redistribution ($\rho_{\rm loc,S}=0.4,\rho_{\rm lat,S}=\rho_{\rm uni,S}=0$); in the presence of only latitudinal redistribution ($\rho_{\rm lat,S}=0.4,\rho_{\rm loc,S}=\rho_{\rm uni,S}=0$); and in the presence of only uniform redistribution  ($ \rho_{\rm uni,S}=0.4, \rho_{\rm loc,S}=\rho_{\rm lat,S}=0$).

\begin{table}[!htb]
\centering
\caption{Parameters for an NN Serpentis-like system based on data in \cite{parsons2010}.}
\label{tab:NNserpensis}
\begin{tabular}{l|c|c}
  Parameter & white dwarf & red dwarf \\ \hline \hline
  atmosphere & blackbody & blackbody \\
  exponent in gravity brightening, $g$ & 1 & 0.32 \\
  polar radius, $R(R_\odot)$ & 0.0211 & 0.147\\
  effective temperature, $T_{\rm eff}(K)$ &  57000 &  3500\\
  masses, $M(M_\odot)$ & 0.535 & 0.111 \\
  fraction of reflection,  $\rho_{\rm refl}$ & 1 & 0.6\\
  synchronicity parameter, $F_{\rm sync}$ & 1 & 1 \\
  fillout factor, $f^{\rm R}$\footnote{The fillout factor follows the definition in \citet[Ch.~3.1.6]{kallrath2009}.} & 0.0472 & 0.773 \\
  limb darkening (LD): \\ \hline
  model & logarithmic & logarithmic \\
  coefficient $x_{\rm LD}$ & 0.5 & 0.5\\  
  coefficient $y_{\rm LD}$ & 0.5 & 0.5\\ 
 
  orbit: \\ \hline
  period, $P({\rm day})$ & \multicolumn{2}{c}{0.1300801714}\\
  eccentricity, $\epsilon$ & \multicolumn{2}{c}{0}\\
  systemic velocity, $\gamma({\rm km/s})$ & \multicolumn{2}{c}{0}\\
  inclination, $\iota(\deg)$  & \multicolumn{2}{c}{89.6} \\
  mass ratio, $M_2/M_1$ & \multicolumn{2}{c}{0.207} \\
  semi-major axis, $a(R_\odot)$ & \multicolumn{2}{c}{0.934}
\end{tabular}
\end{table}

\begin{figure}[!htb]
\centering
\includegraphics[width=14cm]{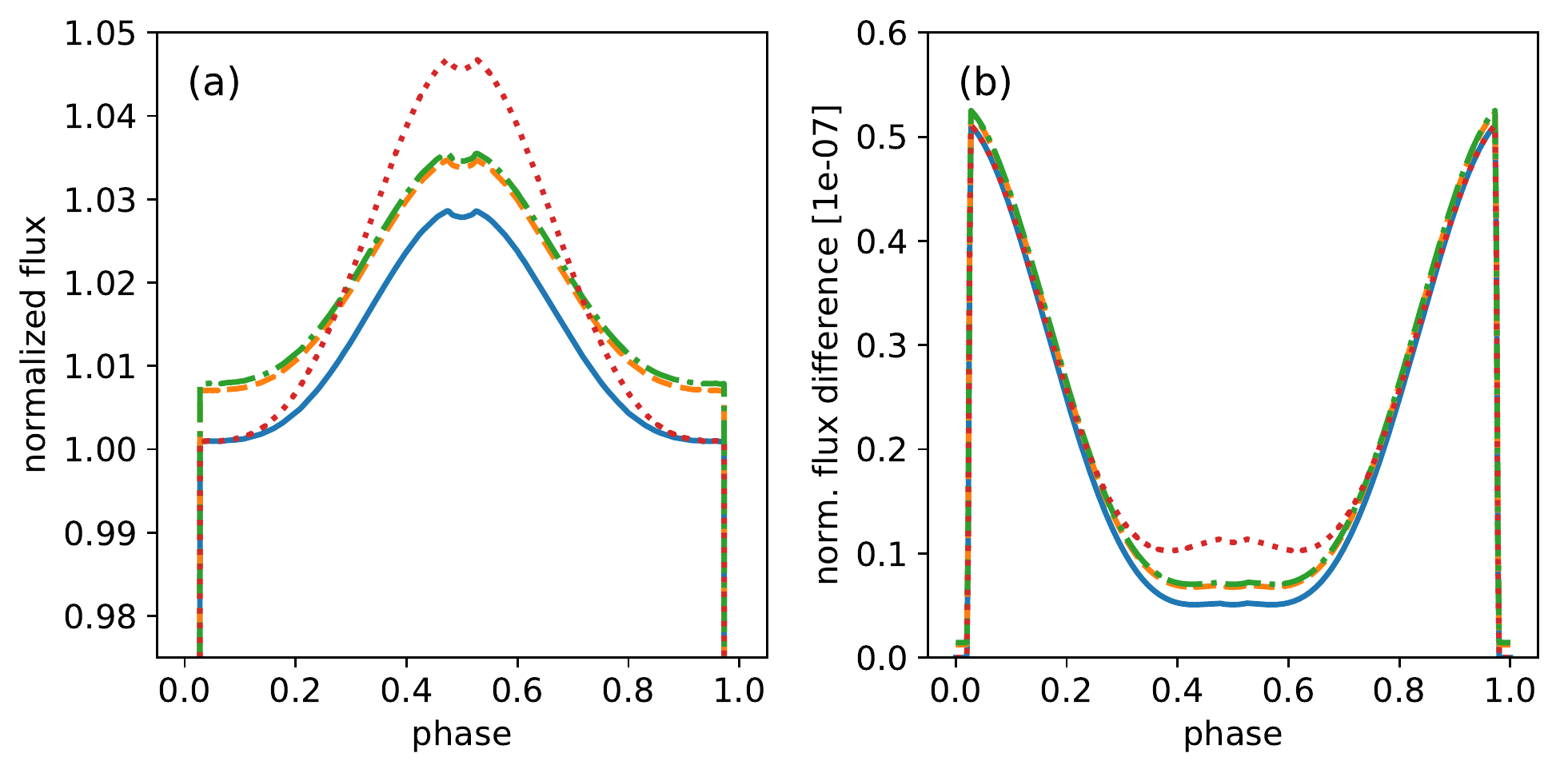}
\includegraphics[width=14cm]{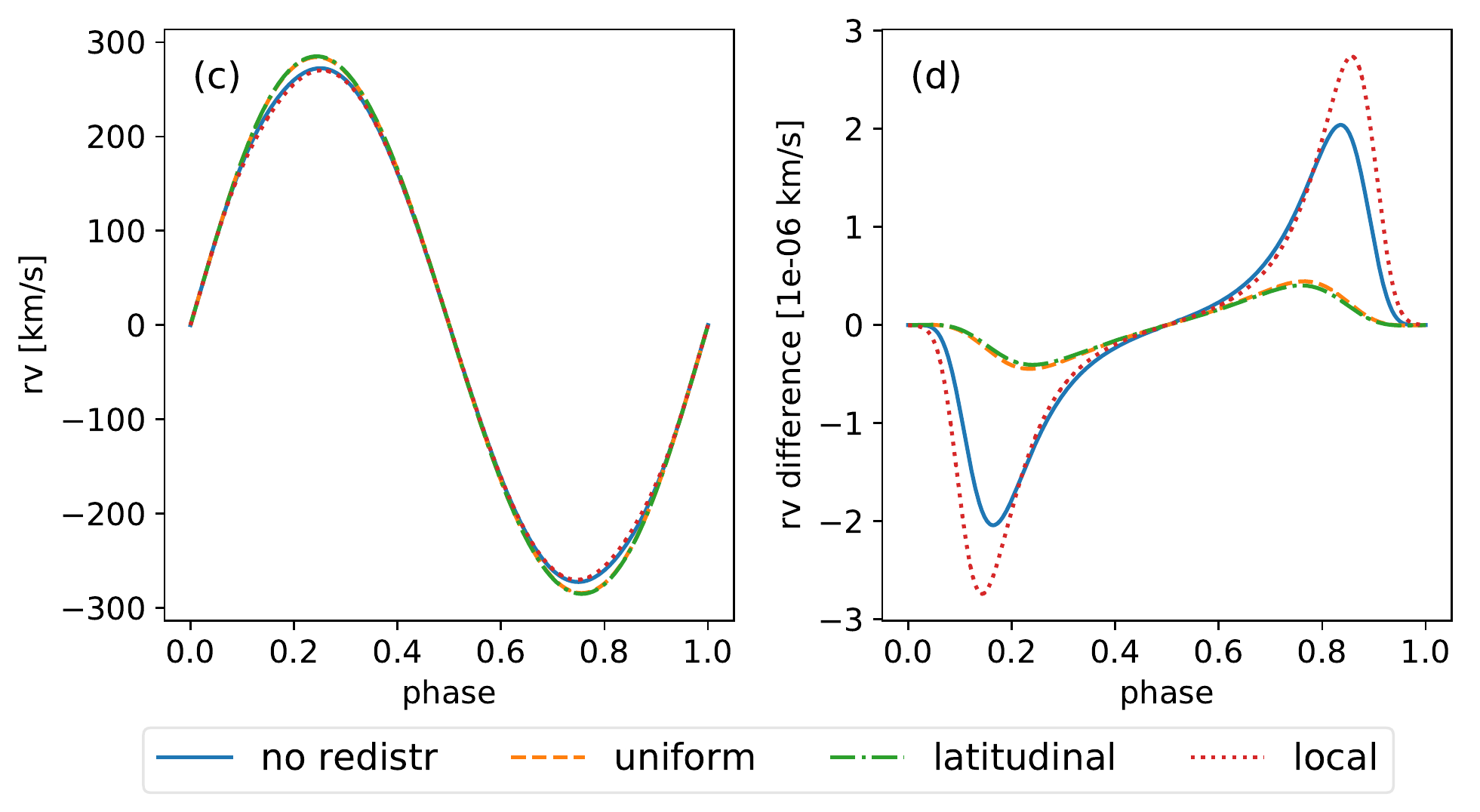}
\caption{The computed light curve (top row) and radial velocity curves (bottom row) of a NN Serpentis-like system using Lambertian reflection with different redistribution schemes (left column) and the difference between the curves obtained using the Lambertian scheme and Wilson's reflection approach (right column).  The models are normalized such that the bolometric luminosity (prior to any irradiation effects) of the primary component is kept fixed at $4 \pi$ between different models, such that, in isolation, it would effectively contribute unity to the overall flux. As the secondary component is much less luminous, the majority of additional flux is from the irradiation on the secondary component by the primary, which differs between these different schemes.}
\label{fig:lc_rv_NNSer}
\end{figure}

We assume that the albedos (i.e., the fraction of the flux that is reflected directly from the surface) of the primary and secondary stars are $\rho_{\rm P}=1$ and $\rho_{\rm S}=0.6$, respectively, and that the radii ($l$; see Eq.~\ref{eq:Glocal}--\ref{eq:Ghoriz})  of the latitudinal and local redistribution processes are given by $l/R= 20^\circ \doteq 0.35$. The bolometric LCs and radial velocity curves of such a system are presented in Fig.~\ref{fig:lc_rv_NNSer}. Note that the calculation of the approximate passband-dependent models would be possible using \citeauthor{wilson1990}'s approach of spectral re-processing of bolometric results, as outlined in Ch. \ref{ch:temp_nonbol}, which we elect to forego on account of clarity and consistency.

The redistribution increases the emitted flux. The strongest increase in comparison to that without redistribution, around 2\%, is noticeable with local redistribution, as we can seen from Fig.~\ref{fig:lc_rv_NNSer}a. This is because it effectively increases the reflectivity of the object, resulting in an increased radiosity/fluxes around the secondary eclipse, when the reflection is at its maximum. The latitudinal and uniform redistribution have a similar effect on the LCs with an approximate 8\% increase from the case without redistribution. Because the uniform redistribution spreads the incoming flux over a wider area (the whole body) than the latitudinal redistribution, the flux of the former case is necessarily smaller than the flux of the latter case. 

The redistribution does not affect radial velocities as strongly as fluxes; see Fig.~\ref{fig:lc_rv_NNSer}c. Interestingly, we obtain a similar radial velocity curves for pairs of latitudinal and uniform redistributions, and local and no redistributions. This is a consequence that the local redistribution affecting only a small area on the surface in comparison to the uniform and latitudinal redistributions. 

Differences between the Wilson and Lambertian reflection schemes in LCs and radial velocities are of the order of magnitude $10^{-7}$ and $10^{-8}$ and presented in Fig.~\ref{fig:lc_rv_NNSer}b and Fig.~\ref{fig:lc_rv_NNSer}c, respectively. Currently, these differences are not measurable in practice.  The fluxes obtained using Wilson's approach are larger than those using Lambertian reflection, which seems to be generally true in a binary configuration. This follows from the fact that the limb-darkened (Wilson) diffusion of light amplifies the intensities in the direction nearly normal to the surface in comparison to Lambertian diffusion, where $D/D_0 > 1/\pi$ for $\mu\approx 1$ for almost all limb darkening coefficients. The intensities in the direction nearly orthogonal to the surface are the most important in the transfer of energy between the bodies, yielding an increase in reflected fluxes and consequently an increase in radiosity. The differences between the radial velocity curves obtained using either the Wilson or Lambertian reflection and some redistribution type are pairwise similar in shape for the local and without redistributions, and the uniform and latitudinal redistributions. The differences of the latter pair are generally smaller, because the redistribution to a wider area has the effect of blurring out local differences in radiosity across the lobe.
 
Following the analysis presented in Sec. \ref{sec:detect_discriminate} we calculate $\sigma_{\rm max}$ and $\sigma_{\rm min}$ by varying theredistribution parameters of the second star for the cases discussed in Fig.~\ref{fig:lc_rv_NNSer} and find that $\sigma_{\rm max}\approx 0.48$, $\sigma_{\rm min}=0.00285$, and the resulting conditional number $\kappa \approx 169$ (Eq. \ref{eq:cond_nr}) This suggests that these cases are far from degenerate, and according to Eq. (\ref{eq:var_C}), the effect of the redistribution in the LC can vary by up to two-orders in magnitude, at fixed redistribution parameters.

\section{Conclusions}

In this paper, we present a general framework for dealing with quasi-stationary irradiation effects between astrophysical bodies. It extends the reflection schemes presented in \citet{prsa2016} to include redistribution, thereby making the framework energy conserving. This framework can essentially be used to describe any arbitrary pattern of redistribution, but here we focus on three possible redistribution processes for nearly spherical bodies, where we can at least partially justify the functional form of redistribution operators. 

We have demonstrated the framework on a toy binary system resembling NN Serpentis in order to confirm that a significant part of the irradiation is absorbed, and therefore that the redistribution effect should be a noticeable. In the considered case, the differences between reflection schemes are very small and not measurable at the current best precision of observational measurements.

As highlighted by \cite{wilson1990}, a ``complete'' treatment of the irradiation effect can be broken into four main components: geometrical, bolometric energy exchange, irradiated stellar atmospheres, and induced changes to envelope structure.  The framework presented here accurately treats the first two parts exactly, and importantly, with the inclusion of true energy conservation\footnote{Some fraction of the energy may contribute to changes in the structure of the irradiated stellar envelope; however, our treatment assumes that this is a quasi-stable effect and thus the energy balance comprises only reflection and redistribution (with no additional fraction altering the envelope structure).}.  The final two aspects are clearly intertwined and highly dependent upon the system parameters, making their accurate treatment computationally expensive and somewhat impractical when attempting to model real-world systems.  As such, the framework presented here represents the most comprehensive treatment of irradiation in binary stars to date in the direction laid out by \cite{wilson1990} and expanded upon by \cite{budaj2011}.

Beyond the clear open questions about the impact of irradiation on the stellar atmospheres and structures (and more generally, on the validity of continuing to use non-irradiated models as representative of irradiated binary stars), there are several aspects of the redistribution and reflection processes that are far from being understood.  Reflection is essentially characterized completely by the bolometric albedo, with theoretical considerations predicting a strong dependence on stellar effective temperature, which has yet to be confirmed by observations \citep[see, e.g.,][]{claret2001}.  Redistribution is even more poorly understood with very few theoretical constraints available.  For example, it is particularly unclear which (if any) of the functions (uniform, local, latitudinal) presented in this work is the most physical way of describing the redistribution process, and furthermore, in close systems where the components are gravitationally distorted, how is the redistribution process affected by such deviations from sphericity.  PHOEBE, combined with the extended framework presented here, currently represents the most complete modeling tool with which to address these open questions observationally.

\acknowledgements

This work was supported by the NSF AAG grant \#1517474, which we gratefully acknowledge.  This research has been supported by the Spanish Ministry of Economy and Competitiveness (MINECO) under the grant AYA2017-83383-P. 


\software{ 
    SHELLSPEC   \citep{budaj2004},
    \phoebe 2.1 \citep{prsa2016},
    astropy     \citep{astropy},
    matplotlib  \citep{matplotlib},
    numpy       \citep{numpy}
}

\bibliographystyle{aasjournal}
\bibliography{refl_redistr}

\appendix

\section{Radiosity equations in integral form} \label{sec:integral_form}

To better illustrate the principles, let us write out the reflection model equations in the integral form for two convex bodies, labeled as A and B, with the corresponding surfaces ${\cal M}_{\rm A}$ and ${\cal M}_{\rm B}$. The irradiation operator can be written as
\begin{equation*}  
  \hat {\cal Q} F({\bf r}) = 
  \int_{\cal M} Q({\bf r}, {\bf r}') F({\bf r}') {\rm d}A({\bf r}')\>.
\end{equation*}
The radiosity equation for \citeauthor{wilson1990}'s model, Eq.~(\ref{eq:wilson_reflection}), for a point ${\bf r}_{\rm B}$ on the surface of star $B$ can be written as
\begin{equation*}
  F_{\rm out} ({\bf r}_{\rm B}) = F_0  ({\bf r}_{\rm B}) + 
  \rho({\bf r}_{\rm B}) \int_{\cal M_{\rm A}}\!\!
  Q({\bf r}_{\rm B}, {\bf r}_{\rm A}) 
  \frac{D(\widehat{({\bf r}_{\rm B} - {\bf r}_{\rm A})}\cdot \hat {\bf n}({\bf r}_{\rm A}))}{D_0 ({\bf r}_{\rm A})} 
  F_{\rm out}({\bf r}_{\rm A})
  {\rm d} A ({\bf r}_{\rm A})
  \>.
\end{equation*}
This is equivalent to the relation given in \cite{wilson1990}, except that it uses the radiosity $F_{\rm out}$ instead of the flux density excess due to reflection, $F_{\rm out}/F_0$.

The irradiance $F_{\rm in}$ for \citeauthor{prsa2016}'s reflection model, Eq.~(\ref{eq:phoebe_reflection}), at ${\bf r}_{\rm B}$ on the surface of star $B$ in integral form is
\begin{equation*}
  F_{\rm in} ({\bf r}_{\rm B}) =
 \int_{\cal M_{\rm A}}\!\! Q({\bf r}_{\rm B}, {\bf r}_{\rm A})
 \left [
  \frac{D(\widehat{({\bf r}_{\rm B} - {\bf r}_{\rm A})}\cdot \hat {\bf n}({\bf r}_{\rm A}))}{D_0 ({\bf r}_{\rm A})}F_{\rm 0}({\bf r}_{\rm A}) +
   \rho({\bf r}_{\rm A})\frac{F_{\rm in}({\bf r}_{\rm A})}{\pi} \right ]
  {\rm d} A ({\bf r}_{\rm A})
   \>.
\end{equation*}

\section{Irradiation approximation in a two-body system} \label{sec:1d_model}

Here we are providing a highly simplified model of irradiation for a binary system. In this model, we reduce the bodies to points with a night and day side and make a rough estimate of the flux exchange on a line (the name one-dimensional comes from here ) between the day sides via irradiation and of redistribution between the day and night sides. The approximation is based on approximating average response of the operator using the mean-field approach.

\subsection{The mean-field approximation}\label{sec:mean_field}

All operators introduced in this paper, i.e reflection $\hat \Pi$, redistribution $\hat {\cal D}$ and radiosity $\hat {\cal L}_*$ ($*={\rm L}, {\rm LD}$) operators, preserve the positivity of the functions, and all functions describing the irradiation process, i.e. irradiance $F_{\rm in}$, radiosity $F_{\rm out}$, and exitances $F_{\rm 0}$ and $F_{\rm 0}'$, are defined on the surfaces of bodies and are non-negative.

In the mean-field approximation approach, we decompose a function $F$ defined on a surface $\cal S$ of area $A$ into its surface average $\langle F \rangle$, defined as
$$
    \langle F \rangle 
    = \frac{1}{A} \int 
      F({\bf r})
      {\rm d} A({\bf r})
$$
and the deviation  $\delta F$ from it, writing $F = \langle F \rangle + \delta F$. Then, an action of some operator $\hat {\cal O}$ on the function $F$ reads
$$
  \hat {\cal O} F = 
  \hat {\cal O}\langle F \rangle +  
  \hat {\cal O} \delta F\>.
$$
Here, for the operators and functions used, the first term on right is the dominant contribution. In the mean-field approximation, we drop the term depending on fluctuations and approximate the surface average of operator image of an function as
$$
 \langle \hat {\cal O} F \rangle \approx 
  \langle F \rangle  \langle \hat {\cal O}\rangle\>
  \qquad{\rm and}\qquad
  \langle \hat {\cal O}\rangle \equiv \langle \hat {\cal O} 1\rangle \>.
$$
We call $\langle \hat {\cal O}\rangle$ the average operator and is defined as the surface average of the image of the constant function equal to 1.

\subsection{One-dimensional model of irradiation} \label{sec:1d_model_details}

Let us discuss a system of two convex bodies, labeled as A and B, with the Lambertian reflection from the surfaces and introduce the intrinsic exitances $F_{0,b}$, updated intrinsic exitances $F_{0,b}'$, exiting $F_{{\rm out},b}$ and entering $F_{{\rm in}, b}$ radiosities defined for both bodies $b = {\rm A}, {\rm B}$. The updated intrinsic exitances $F_{0,b}'$ and exiting radiosities $F_{{\rm out},b}$ are expressed as
\begin{equation}
   F_{0,b}' = F_{0,b} +  \hat{\cal D}_b (1 - \hat \Pi_b) F_{{\rm in},b} \quad {\rm and} \quad
   F_{{\rm out},b} = F_{0,b}' +  \hat \Pi_b F_{{\rm in},b}  \>,
\end{equation}
where $\hat{\cal D}_b$ and $\hat \Pi_b$ are the redistribution operator and the reflection operators associated with the body $b$, respectively. Consequently, we can separate the irradiation Eq. (\ref{eq:lambert_irrad}) into a system of two equations, each dealing with the irradiation of the considered star:  
\begin{align}
  F_{\rm in,B} &= 
  \hat {\cal L}_{\rm LD, A\to B} F_{\rm 0,A} + 
  \left[
    \hat{\cal L}_{\rm LD, A\to B} \hat {\cal D}_{\rm A} (\mathbb{1} - \hat\Pi_{\rm A})  +  
    \hat{\cal L}_{\rm L, A\to B} \hat\Pi_{\rm A}
  \right] F_{\rm in, A}\>, \label{eq:two_body_1}\\
  F_{\rm in,A} &= 
  \hat {\cal L}_{\rm LD, B\to A} F_{\rm 0,B} + 
  \left[
    \hat{\cal L}_{\rm LD, B\to A} \hat{\cal D}_{\rm B} (\mathbb{1} - \hat\Pi_{\rm B})  +  
    \hat{\cal L}_{\rm L, B\to A} \hat\Pi_{\rm B}
  \right] F_{\rm in, B} \label{eq:two_body_2}\>.
\end{align}
The radiosity operators $\hat{\cal L}_{{\rm LD}, b\to b'}$ and $\hat{\cal L}_{{\rm L}, b\to b'}$ describe the limb-darkened and Lambertian diffusion of light from the surface of body $b$ onto the surface of body $b'$. For the sake of simplicity, we assume a constant reflection on both bodies ($b={\rm A}, {\rm B}$): 
$$
  \hat \Pi_b = \rho_b \mathbb{1}, \qquad \rho_b = {\rm const}\>.
$$

The surface of the bodies can be divided into an illuminated part, called day side, and a non-illuminated part, called night side. The irradiation equations (\ref{eq:two_body_1}) - (\ref{eq:two_body_2}) describe processes only on the day side, and consequently, $F_{{\rm in}, b}$ is non-zero only on that side of body $b ={\rm A} ,{\rm B}$. Additionally, we may notice that
\begin{align}
  F_{{\rm out},b}({\bf r}) &= 
  F_{0,b}'({\bf r}) + \hat \Pi_b F_{{\rm in}, b}({\bf r})
  &{\bf r} \in \textrm{day side}\>, \\
  F_{{\rm out}, b}({\bf r}) &=  
   F_{0,b}'({\bf r})
  &{\bf r} \in \textrm{night side}\>.
  \label{eq:Fout_nightday}
\end{align}

Next, we introduce the averages over the whole surface $\langle \cdot \rangle$, over the day side $\langle \cdot \rangle_{\rm day}$, and over the night side $\langle \cdot \rangle_{\rm night}$ of a considered body. The areas of the entire surface and of the day side of body $b$ are denoted by $A_b$ and $A_{b,{\rm day}}$, respectively. Taking into account Eq. (\ref{eq:Fout_nightday}), we can conclude for body $b$ that
\begin{equation}
  \langle F_{{\rm out}, b}\rangle =
 p \langle F_{{\rm out}, b}\rangle_{\rm day}  + (1-p)\langle F_{0, b}'\rangle_{\rm night} \>,
\end{equation}
where we use the ratio of the areas $p=A_{b,{\rm day}}/A_b$.

We start the approximation of the irradiation model by decomposing all involved quantities, i.e., $F_{0,b}$, $F_{0,b}'$ and  $F_{{\rm out}, b}$, into their day- and night-side counterparts. Then, we take the surface average of the irradiation Eqs. (\ref{eq:two_body_1})-(\ref{eq:two_body_2}) and the quantities over the both sides of the bodies separately. The irradiation equations are only defined on the day sides, and so the night side averages yield zero. We approximate the actions of the operators using a mean-field approach, whereby we replace functions defined across the day and night sides with their corresponding average; see Appendix \ref{sec:mean_field} for details. Following this idea, we approximate the averages of the operators acting on a function $F$ defined over the day side of body $b$ according to the next rules:
\begin{align}
   \langle \hat {\cal L}_{*, b\to b'} F \rangle_{\rm day} &\approx L_{*,b\to b'} \langle F \rangle_{\rm day}\>,\\
   \langle \hat {\cal D}_b F \rangle_{\rm day} &\approx \eta_b \langle F \rangle_{\rm day} \>,\\
   \langle \hat {\cal D}_b F \rangle_{\rm night} &\approx (1-\eta_b) \langle F \rangle_{\rm day} \>, 
\end{align}
where $* ={\rm LD}, {\rm L}$ is labeling different surface behaviors. Here we introduce the model constants $L_{*,b\to b'}$ and $\eta_b$ that describe the effective action of the radiosity and redistribution operators on the surface-averaged incoming radiosity $\langle F_{\rm in,b}\rangle_{\rm day}$. More precisely, $L_{*,b\to b'}$ represents the average ratio of emitted energy transferred from points on body $b$ to body $b'$, and $\eta_b$ quantifies the average ratio of absorbed energy that is re-emitted on the same side.

\begin{figure}[!htb]
    \centering
    \includegraphics[width=10cm]{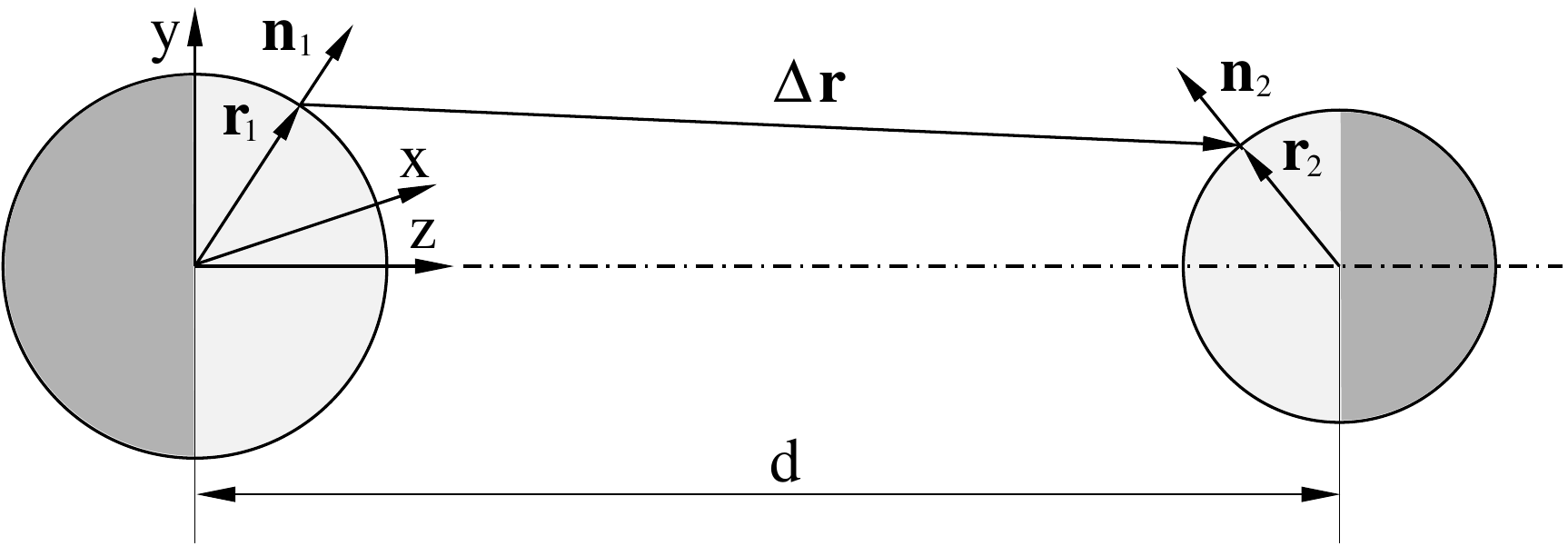}
    \caption{Scheme of two spheres used in the 1D irradiation model.}
    \label{fig:spheres_model}
\end{figure}

\begin{figure}[!htb]
    \centering
     \includegraphics[width=7.5cm]{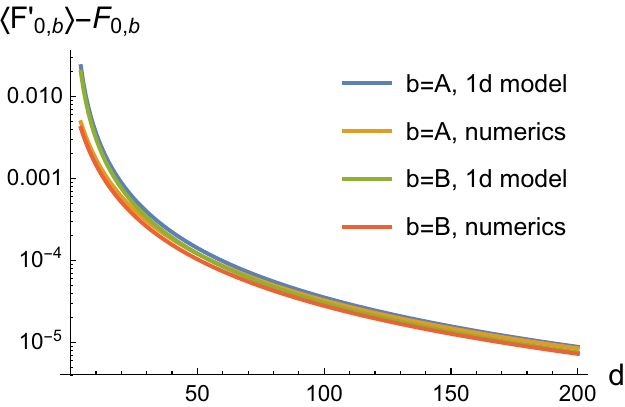}
    \includegraphics[width=7.5cm]{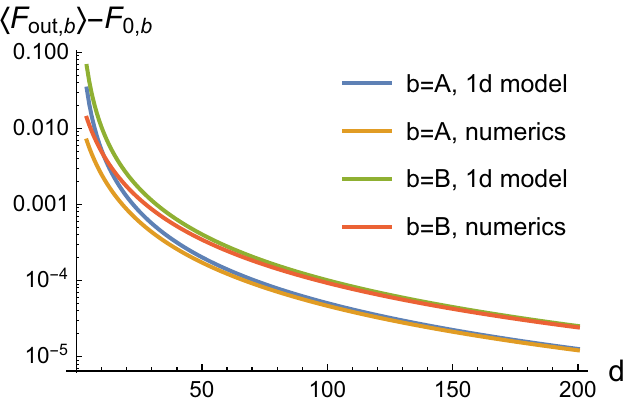}
    \caption{Comparing the average updated exitance $\langle F_{0,b}' \rangle$ (left) and average radiosity $\langle F_{{\rm out},b} \rangle$ (right) obtained from numerical calculations using 10k+ triangles with linear limb darkening ($D(\mu) = 1 - x(1-\mu)$ with $x=0.3$) and a one-dimensional model using global redistribution at parameters $r_{\rm A}=2$, $r_{\rm B}=1$, $\eta_{\rm A} = \eta_{\rm B} = 1/2$, $F_{0,{\rm A}}=1$, $F_{0,{\rm B}}=2$, $\rho_{\rm A}=0.3$, and $\rho_{\rm B}=0.7$.}
    \label{fig:1dmodel_res}
\end{figure}

The averages of the quantities describing the irradiation process over the day and night sides are then given by
\begin{align}
  \langle F_{0,b}' \rangle_{\rm day} &=
  \langle F_{0,b} \rangle_{\rm day} +
 (1-\rho_b) \eta_b \langle F_{{\rm in},b} \rangle_{\rm day}\>, \\
  \langle F_{0,b}' \rangle_{\rm night} &=
  \langle F_{0,b} \rangle_{\rm night} +
  (1-\rho_b) (1-\eta_b)\langle F_{{\rm in},b} \rangle_{\rm day}\>,\\
  \langle F_{{\rm out},b} \rangle_{\rm day} &=
  \langle F_{0,b}' \rangle_{\rm day} +
  \rho_b \langle F_{{\rm in},b} \rangle_{\rm day}\>,\\
  \langle F_{{\rm out},b} \rangle_{\rm night} &=
  \langle F_{0,b}' \rangle_{\rm night}\>.
\end{align}
By introducing additional auxiliary coefficients,
\begin{align}
T_{b\to b'} &= (1- \rho_b)  \eta_b L_{{\rm LD}, b\to b'}  +  \rho_b L_{{\rm L},b\to b'}\>,\\
G_{b'} &= L_{{\rm LD}, b\to b'} \langle F_{0,b}\rangle_{\rm day}\>,
\end{align}
the average of the irradiation Eqs. (\ref{eq:two_body_1})-(\ref{eq:two_body_2}) can be rewritten into a simple system of two scalar equations involving variables $\langle F_{{\rm in}, b} \rangle_{\rm day}$ for body $b={\rm A}, {\rm B}$:
\begin{align}
   \langle F_{\rm in, A} \rangle_{\rm day} &= 
   G_{\rm A} + T_{\rm B\to A} \langle F_{\rm in, B} \rangle_{\rm day}\>,\\
    \langle F_{\rm in, B} \rangle_{\rm day} &= 
    G_{\rm B} + T_{\rm B\to A} \langle F_{\rm in, B} \rangle_{\rm day}\>.
\end{align}
The solution of this system is equal to
\begin{equation}
 \left[ 
 \begin{array}{cc}
   \langle F_{\rm in, A} \rangle_{\rm day} \\
   \langle F_{\rm in, B} \rangle_{\rm day}
 \end{array}
 \right] =  
 \frac{1}{1 - T_{\rm B\to A }T_{\rm A\to B}}\left[
 \begin{array}{cc}
   G_{\rm A} + T_{\rm B\to A} G_{\rm B} \\
   T_{\rm A\to B} G_{\rm A} + G_{\rm B}
 \end{array} \right]\>.
 \label{eq:1dmodel_results}
\end{equation}
Let us now assume that the bodies are perfect spheres of radii $r_{\rm A}$ and $r_{\rm B}$ and their centers separated by distance $d$, as depicted in Fig. \ref{fig:spheres_model}. We set the intrinsic exitance $F_{0,b}$ to be constant over the surface of each star separately.
We are interested in the limit $r_b \ll d$ in which we can approximate model constants as
\begin{align}
    L_{*, b\to b'} &\approx \frac{r_b^2}{2 d^2} \>, \\
    \eta_b &\approx 
    \left \{
        \begin{array}{lll}
        \frac{1}{2} &:&  \textrm{global or latitudinal redistribution}\\
        1 &:& {\rm local}
        \end{array}
    \right . \>.
\end{align}
We find numerically that the coefficient $L_{*,b\to b'}$ is independent of the type of energy diffusion, labeled by $*$. An explicit formula for the coefficient is given in Appendix \ref{sec:1d_coef}. In the following, we compare the average exiting radiosity  $\langle F_{{\rm out}, b}\rangle$ and average updated exitance $\langle F_{0, b}'\rangle$ as functions of the distance $d$ between the bodies obtained in a one-dimensional model, given by
\begin{align}
 \langle F_{{\rm out}, b}\rangle &=  
 F_{0, b} + \frac{1}{2}\langle F_{{\rm in},b}\rangle_{\rm day} \>, \\
 \langle F_{0, b}'\rangle &=  
 F_{0, b} + 
 \frac{1}{2}(1-\rho_b)\langle F_{{\rm in},b}\rangle_{\rm day}\>,
\end{align}
and that obtained from numerical calculations by discretizing the surface into triangles, as described in Sec. \ref{sec:discret}. In the considered limit, we may take the area of the day and night sides to be identical; for details, see Appendix \ref{sec:1d_coef}. The results are depicted in Fig. \ref{fig:1dmodel_res} and show a good qualitative agreement between the two approaches, especially in the limit of larger separations between bodies. 

\subsection{Coefficients in a 1D model for two-spheres system}\label{sec:1d_coef}

Let us consider a binary system composed of two spheres, labeled by A and B, with radii $r_{\rm A}$ and $r_{\rm B}$, and their centers separated by a distance $d$, as depicted in Fig. \ref{fig:spheres_model}. The area of the illuminated side of the sphere $b$ is given by 
\begin{equation}    
    A_{b,{\rm day}} = 
    2\pi r_b^2 \left(1 - \frac{r_b' - r_b}{d}\right)\>.
\end{equation}
We assume that the limb darkening law is constant across the surface. The coefficients $L_{*, b\to b'}$ with $*={\rm L}, {\rm LD}$ introduced in Appendix \ref{sec:1d_model_details} describing the effect of the radiosity operator on the average incoming radiosity can be identified as the averages of radiosity operator, based on previuos subsection:
\begin{equation}
   \left\langle {\hat{\cal L}}_{*, b\to b'}\right\rangle_{\rm day} \equiv L_{*, b\to b'}\>.
\end{equation}
In the considered case, this coefficient can be expressed as an integral over all possible pairs of points on the two spheres with their line of sight unobstructed:
\begin{equation}
  \begin{split}
    L_{*, b\to b'}
    =&
    \frac{(r_b r_{b'})^2}{A_{b',{\rm day}}}
    \int {\rm d} \Omega(\hat {\bf n}_b)
    \int {\rm d} \Omega(\hat {\bf n}_{b'})\,
    U(\Delta {\bf r}  \cdot \hat {\bf n}_b)
    U(\Delta {\bf r}  \cdot \hat {\bf n}_{b'})\\
    &\cdot\frac{
        (\Delta {\bf r}  \cdot \hat {\bf n}_b)
        (\Delta {\bf r}  \cdot \hat {\bf n}_{b'})
    }
    {\|\Delta {\bf r}\|^4}
    \frac
        {D_*(\widehat{\Delta {\bf r}}\cdot \hat {\bf n}_b)}
        {D_{*,0}}\>,
    \end{split}
\end{equation}
where integrations are carried out over full solid angles, and we should remind ourselves that $D_*$ is the limb darkening function and $D_{*,0}$ is its integral over a hemisphere:
\begin{equation}    
 D_{*,0} = 2\pi \int_0^1 {\rm d} \mu\> D_*(\mu) \mu \>.
\end{equation}
Here we use the unit-step function $U(x) = \{1\mathpunct{:}~ x \ge 0; 0\mathpunct{:}~ {\rm otherwise} \}$, and the vector connecting the pairs of points is equal to
\begin{equation}
\Delta {\bf r}  = 
r_2 \hat {\bf n}_2 + d \hat{\bf k} - r_1 \hat {\bf n}_1\>.
\end{equation}
We find numerically that the leading order of the average operator in the limit $d\to\infty$ behaves as
$$
  L_{*, b\to b'} \sim \frac{r_b^2}{2 d^2} \>,
$$
and this behavior seems to be independent of the chosen limb darkening law labeled by the index $*$.

\end{document}